\documentclass[aps,prx,superscriptaddress,floatfix,twocolumn,longbibliography]{revtex4-2}

\usepackage{amsmath,dcolumn,bm,amsthm,tabularx,subfigure}
\usepackage[colorlinks=true,urlcolor=blue,citecolor=blue,linkcolor=blue]{hyperref}
\usepackage{mathrsfs,times}
\usepackage{bbold,dsfont}
\usepackage{graphicx,graphics,color}
\usepackage{mathtools,nicefrac}
\usepackage{slashed}
\usepackage{amsfonts,amssymb}
\usepackage{tikz}
\usetikzlibrary{positioning}
\allowdisplaybreaks

\newcommand{\beq}{\begin{equation}}
\newcommand{\eeq}{\end{equation}}
\newcommand{\bea}{\begin{eqnarray}}
\newcommand{\eea}{\end{eqnarray}}
\newcommand{\widebar}{\overline}

\newcommand{\X}{\widetilde{X}}
\newcommand{\Z}{\widetilde{Z}}
\newcommand{\ZZ}{{Z\!Z}}
\newcommand{\ZX}{{Z\!X}}
\newcommand{\XX}{{X\!X}}

\begin{document}

\title{Lieb-Robinson bound for constrained many-body localization}

\author{Chun~Chen}
\email[Corresponding author.\\]{chunchen@sjtu.edu.cn}
\affiliation{School of Physics and Astronomy, Key Laboratory of Artificial Structures and Quantum Control (Ministry of Education), Shenyang National Laboratory for Materials Science, Shanghai Jiao Tong University, Shanghai 200240, China}

\author{Xiaoqun~Wang}
\email[Corresponding author.\\]{xiaoqunwang@sjtu.edu.cn}
\affiliation{School of Physics and Astronomy, Key Laboratory of Artificial Structures and Quantum Control (Ministry of Education), Shenyang National Laboratory for Materials Science, Shanghai Jiao Tong University, Shanghai 200240, China}
\affiliation{Tsung-Dao Lee Institute, Shanghai Jiao Tong University, Shanghai 200240, China}
\affiliation{Beijing Computational Science Research Center, Beijing 100084, China}

\author{Yan~Chen}
\email[Corresponding author.\\]{yanchen99@fudan.edu.cn}
\affiliation{Department of Physics and State Key Laboratory of Surface Physics, Fudan University, Shanghai 200433, China}
\affiliation{Collaborative Innovation Center of Advanced Microstructures, Nanjing University, Nanjing 210093, China}

\date{\today}

\begin{abstract}

Study how quantum information propagates through spacetime manifold provides a means of identifying, distinguishing, and classifying novel phases of matter fertilized by many-body effects in strongly interacting systems in and out of equilibrium. Via a fuller characterization of key aspects regarding dynamic behaviors of information, we perform such an analysis on constrained many-body localization---a newly proposed fully localized state under infinite-interaction limit---in quasirandom Rydberg blockade spin chain models using thermal out-of-time-order commutators (OTOCs). The OTOC light cones predict a hitherto unknown Lieb-Robinson bound for constrained many-body localization, which is qualitatively different from that of unconstrained many-body Anderson insulators stabilized at weak-interaction limit. Our corroborated numeric and analytic study suggests that constrained many-body localization is a distinct dynamical eigenstate phase whose nonergodicity is beyond local-integral-of-motion phenomenology. Together, these findings consolidate the hierarchy of unconventional quantum dynamics encompassing constrained, unconstrained, and diagonal many-body-localized regimes.     

\end{abstract}

\maketitle

\noindent \textbf{INTRODUCTION}

\noindent Anderson localization (AL) describes absence of diffusion arising from strong disorders \cite{Anderson}. Within last decade, while still under debate, physicists progressively approach a consensus that in one dimension ($1$D), extended version of this noninteracting phenomenon, termed unconstrained many-body localization (uMBL), may survive, provided that the added interparticle interactions are small \cite{Basko,Gornyi,Oganesyan,Abanin,DeRoeck,SuntajsPRB,Panda_2020,SierantThoulessTime,Abanin2021}.

When crafting uMBL theoretical framework, an enduring impetus stems from the quest: what are the fundamental differences between uMBL and its relative AL? Indeed, from a traditional perspective, these phases look similar: (i) they both exhibit insulating behaviors with vanishing dc conductivities at finite temperatures for particle and energy transports. Moreover, (ii) level-spacing statistics based on their respective eigenspectra also fulfil identical Poissonian distribution, suggesting robust emergent integrability that violates ergodicity. It is therefore a milestone (iii) when logarithmic entanglement growth was discovered in uMBL, which markedly contrasts to rapid saturation and sheer absence of entanglement dynamics in AL \cite{Znidaric,BardarsonPollmannMoore}. Soon afterwards, (iv) such dynamical distinctions were magnified through the peculiar OTOC structures of uMBL and the nontriviality of its associated Lieb-Robinson (LR) bound, whereas for AL, it is consistently found that information propagation therein is halted \cite{Fan,HuangZhangChen}. Remarkably, all above-listed four hallmarks, which comprise an assembly of universal properties that defines randomness-induced uMBL, can be interpreted if a full set of local integrals of motion (LIOMs) is assumed \cite{HuseLIOM,Serbyn,Ros}.

Since Anderson's prediction on localization \cite{Anderson}, during last half century, most research efforts were devoted to single-particle localization or the stability of its many-body generalization, especially, for recent $15$ years \cite{Basko,Gornyi}. In view of the mounting relevance of localization to future quantum technology and theory of statistical physics, can there be fully localized phases and mechanisms that are neither AL nor uMBL?   

Inspired by breakthrough in Rydberg experiments \cite{Bernien}, this question was attempted in \cite{Chen} via linking it to circumstances featured by infinite many-body interactions. Specifically, quench disorder was introduced into Rydberg blockade model where two adjacent atoms cannot be simultaneously excited. Particularly, we wondered about if localization persists still in such strongly interacting limits when disorders conflict with interactions via the off-diagonal channel. Despite a tentative tendency toward localization, owing to small system-size flow in exact diagonalization and proximity to nearby transition, finite-size fluctuations combined with Griffiths effect render it unclear whether localization or an intermediate regime is reached, thereby hindering a study on this topic.

To mitigate the issue, in a second paper \cite{ChenChen} of this series we substitute quasiperiodic modulation for quench disorder to bypass Griffiths region when randomizing Rydberg array and show for the first time that robust localization is truly stabilized when particle-particle interaction is infinite in strength and transverse in direction to randomness orientation. Crucially, the ensuing off-diagonal constrained many-body localization (cMBL) is no longer a many-body Anderson insulator.

Concretely, by performing static spectral analyses, we first prove theoretically the full localization via demonstrating (1) Poisson level statistics and (2) strictly zero dc energy conductivity for cMBL. The next step forwarded by \cite{ChenChen} concerns the numerical detection of qualitatively different dynamic behaviors between cMBL and uMBL: (3) contrary to ubiquitous single-log entanglement build-up in canonical uMBL, half-chain von Neumann entanglement entropy in cMBL grows double-logarithmically over time. (4) After parsing in detail the spatial distribution of energy transport, a confined core structure is identified in integrals of motion (IOMs) of cMBL where appreciable nonlocal correlations dominate. These unusual characters [(3) and (4)], which were never seen before, cast doubt on LIOM phenomenology in accounting for cMBL. We therefore report a fully localized phase in \cite{ChenChen} featuring fundamental distinctions from both uMBL and AL.

A closer inspection of cMBL, however, leaves an unresolved impression that on the one hand, the unusual double-log entanglement growth manifests an ultraslow dynamical characteristic of cMBL. While, on the other hand, the considerable confined nonlocal effect embedded in IOMs of cMBL points toward the opposite tendency of a probably faster scrambling. How can these two seemingly conflicting properties be unified in one single cMBL phase?

More importantly, what trait is the built-in genesis that distinguishes cMBL? It is known that albeit slowly, for generic MBL systems, the propagation of quantum information is strictly unbounded \cite{BardarsonPollmannMoore}. Instead, the primary limitation on the efficiency of information transmission is governed by LR bounds \cite{LiebRobinson}, which, serving as the effective \lq\lq speed of light'' for nonrelativistic quantum dynamics, quantify system's locality and causality structures through monitoring noncommutativity between two disjoint operators under unitary time evolution. In absence of constraints, this information flow gives rise to a characteristic logarithmic light cone that designates territory of uMBL phase. What does the light cone of newfound cMBL look like? What is its associated LR bound, and how does it shape cMBL's light-cone front?

It turns out that crux of these questions lies in dynamics of quantum information, i.e., the spatiotemporal information spreading. One proper tool capable of capturing this spacetime complexity beyond autocorrelation function is OTOC, originally proposed by Larkin and Ovchinnikov in semiclassical superconducting theory \cite{LarkinOvchinnikov} and popularized recently by Kitaev in quantum chaos and thermalization \cite{Kitaev},
\beq
C^{{A}{B}}_{\beta}(i,j;t,t')=\frac{1}{2}\langle [{A}(i,t),{B}(j,t')]^\dagger [{A}(i,t),{B}(j,t')] \rangle_{\beta},
\label{otoc_def}
\eeq
where ${A},{B}$ are specified Heisenberg operators, e.g., ${A}(i,t)\coloneqq e^{iHt}{A}(i,0)e^{-iHt}$. See methods section for additional derivations on this definition. In present work, we exclusively focus on thermal ensemble averages of the quantity at infinite temperature, i.e., $\beta=0$, and keep setting $\hbar=1$.

Let's outline the tactic. The central object under scrutiny is OTOC, which amounts to Frobenius norm of commutator squared, thereby linking directly to LR bound. Furthermore, by compiling time evolution of OTOC at fine distances between ${A}$ and ${B}$, one visualizes the spatiotemporal arrangement of the light cone to decode how information is processed. Previous studies \cite{HuangZhangChen} demonstrated that this OTOC light cone is dictated by LR bound. Hence, OTOC comprises a device to bridge the respective propagations of information along space and time under guiding principles of locality and quantum mechanics. Put differently, via computing OTOC light cone, one deduces detailed form of LR bound; the gained knowledge facilitates the deciphering of the complicated OTOC structures.

Practically, OTOC and LR bound assume a vital role in elucidating the subtle underpinnings of unconventional localization in constrained settings. Being a joint spacetime identity, they are bound to yield more comprehensive information about the underlying quantum dynamics than time-dependent half-chain entanglement entropy. For instance, as shown below, a descendent entropy bound compatible with the observed double-log entanglement growth \cite{ChenChen} is derivable from cMBL's LR bound via OTOC-R\'enyi-entropy theorem \cite{Fan,Hosur}. 

\vbox{} \noindent \textbf{RESULTS}

\noindent \textbf{Model}

\noindent The quasirandom Rydberg chain is describable by the Hamiltonian \cite{Chen,ChenChen},
\begin{align}
H_{\textrm{qp}}&=\sum_i [g_i\widetilde{X}_i+h_i\widetilde{Z}_i], \label{constrham}
\end{align}
where $g_i=g_x+W_x\cos[\frac{2\pi i}{\phi}+\phi_x],\ h_i=W_z\cos[\frac{2\pi i}{\phi}+\phi_z]$, and $\X_i\!\coloneqq\!P\sigma^x_i P,\ \Z_i\!\coloneqq\!P\sigma^z_i P$ denote projected Pauli matrices under the global projection $P\coloneqq\prod_i[(3+\sigma^z_i+\sigma^z_{i+1}-\sigma^z_i\sigma^z_{i+1})/4]$, which annihilates motifs of $\downarrow\downarrow$-configuration over any adjacent sites of the chain. Quasiperiodic modulation is then controlled by inverse golden ratio $1/\phi=(\sqrt{5}-1)/2$ and $\phi_x,\phi_z\in[-\pi,\pi)$ are independent random overall phase shifts. Throughout this paper, $W_x\!=\!1$ sets the energy scale, viz., the chain is quasirandom at least along $x$ direction.

{\it Hard-core boson representation.}---To unravel the interplay between finite tunable randomness and infinite interparticle interaction as encapsulated in model (\ref{constrham}), we introduce hard-core boson operators $b_i^\dagger,b_i$ at each site $i$ to describe the local pseudospin-$1/2$ subsystem that mimics the lattice gas of Rydberg atoms with ground state $|g\rangle_i=|\!\!\uparrow\rangle_i$ and excitation state $|r\rangle_i=|\!\!\downarrow\rangle_i$. Formally,
\begin{gather}
b_i^\dagger+b_i=|r\rangle_i\langle g|+|g\rangle_i\langle r|=|\!\downarrow\rangle_i\langle\uparrow\!|+|\!\uparrow\rangle_i\langle\downarrow\!|=\sigma_i^x, \label{boson_sigmax} \\[0.5em]
b_i^\dagger b_i=n_i=|r\rangle_i\langle r|=|\!\downarrow\rangle_i\langle\downarrow\!|=(1-\sigma_i^z)/2. \label{boson_sigmaz}
\end{gather}
Equipped with above expressions, Hamiltonian (\ref{constrham}) can alternatively be mapped onto an array of neutral atoms loaded in Rydberg blockade regime,
\beq
H_{\textrm{qp}}=\sum_i[g_i(b^\dagger_i+b_i)+h_i(1-2n_i)+V_1 n_i n_{i+1}]. \label{constrhamboson}
\eeq
Here $g_i,h_i$ are respectively proportional to onsite Rabi frequency and frequency detuning. Long-range repulsive van der Waals interaction is truncated in (\ref{constrhamboson}) to retain merely nearest-neighbor interaction whose strength $V_1$ is lifted to infinity, producing a blockade radius of $a<R_b<2a$.

{\it Mixed-field Ising representation.}---Since our attention is focused on a single lattice Hamiltonian, it is beneficial to explain why the pursued physics does not suffer drawbacks of being specialized. Note that in terms of spin operators from (\ref{boson_sigmax}) and (\ref{boson_sigmaz}), Hamiltonian (\ref{constrhamboson}) can also be recast into the infinitely interacting version of the paradigmatic mixed-field Ising chain,
\beq
H_{\textrm{qp}}=\sum_i[g_i\sigma^x_i+h_i\sigma^z_i+\frac{V_1}{4}(1-\sigma^z_i-\sigma^z_{i+1}+\sigma^z_i\sigma^z_{i+1})].
\eeq
Considering that a grand portion of uMBL theoretical foundation is framed upon Imbrie's quasiexact mathematical proof of many-body-generalized Anderson insulator in weakly interacting but strongly disordered mixed-field Ising chain, there is a good reason to envision that current series of works targeting the same classic model bears the originality, generality, and significance to stimulate continued efforts on this emerging frontier of unconventional MBL.

The kinetic constraint has been implemented in Rydberg blockade chain \cite{Bernien} and the quasiperiodic modulation has become feasible in experiments \cite{SchreiberBloch,LukinGreiner} to achieve the signature of MBL under the unconstrained circumstances. Accordingly, in addition to pure theoretical interests, the actual merit of model (\ref{constrham}) resides in its high experimental pertinence.

\vbox{} \noindent \textbf{LR bound in cMBL}

\noindent Upon preparation so far, we now declare our major analytic and numeric findings regarding the dynamics of quantum information in randomized Rydberg blockade chains.

To proceed, we first state an assertion on the analytic expression of LR bound for cMBL and contrast this postulate and its ensuing light cone with established results of uMBL. Numeric evidence based on exact diagonalization is next supplied to verify the postulated bound formula via a demonstration on its correctness in capturing OTOC profiles along both space and time directions. Finally, we show that LIOM phenomenology, despite powerful enough to account for uMBL's LR bound, fails in cMBL. As pinpointed by \cite{ChenChen}, the missing piece stems from thermal-like core structures embedded in IOMs of cMBL. To incorporate this key ingredient, a new phenomenological theory is put forward to synthesize multiple intrinsic properties of constrained localization, thereby allowing a qualitative understanding about its unusual LR bound.

{\em Postulate.}---One principal result of present work is to propose the following new LR bound for the definition of cMBL: 
\beq
\|[A(x,t),B(0,0)]\|_{\textrm{cMBL}}\lessapprox c \exp[-\eta\ln|x|+\xi\ln\ln|t|],
\label{LR_cMBL}
\eeq
where the positive exponents satisfy $\xi>\eta$ and $\|\cdot\|$ stands for the operator norm, i.e., the modulus of operator's maximal singular value, which exceeds operator's Frobenius norm, $\|A\|_{\textrm{F}}\coloneqq\sqrt{\textrm{Tr}(A^\dagger A)/\textrm{Tr}(\mathds{1})}\leqslant\|A\|$.

From (\ref{LR_cMBL}), it is recognized that development of OTOC in cMBL is featured concurrently by a logarithmic growth over time (up to some power) and a power-law fall-off across space. Select a threshold $\varepsilon$, the cMBL's OTOC light-cone front can then be captured by
\begin{equation}
\ln|t|\approx(\varepsilon/c^2)^{1/(2\xi)}\cdot|x|^{\eta/\xi}.
\label{LC_cMBL}
\end{equation}

Being a comparison, we list below the known LR bound for the familiar uMBL \cite{KimChandranAbanin,HuangZhangChen}:
\beq
\|[A(x,t),B(0,0)]\|_{\textrm{uMBL}}\lessapprox f \exp[-\varsigma|x|+\nu\ln|t|].
\label{LR_uMBL}
\eeq

As opposed to (\ref{LR_cMBL}), Eq.~(\ref{LR_uMBL}) predicts that for uMBL, the rise of OTOC follows a power-law function of time, meanwhile its weight along the spatial coordinate becomes exponentially attenuated when deviating from the light cone. The resulting front in this case is given by
\begin{equation}
\ln|t|\approx\frac{1}{2\nu}\ln(\varepsilon/f^2)+\frac{\varsigma}{\nu}|x|.
\label{LC_uMBL}
\end{equation}

Be aware that this seemingly minor quantitative change in the exponent of light-cone front [from $1$ in (\ref{LC_uMBL}) to $\eta/\xi$ in (\ref{LC_cMBL})] results from the qualitative difference between the two distinct categories of LR bounds in their actual contents of the respective formulae [see (\ref{LR_cMBL}) and (\ref{LR_uMBL})].

\begin{figure*}[tb]
\centering
\includegraphics[width=0.85\linewidth]{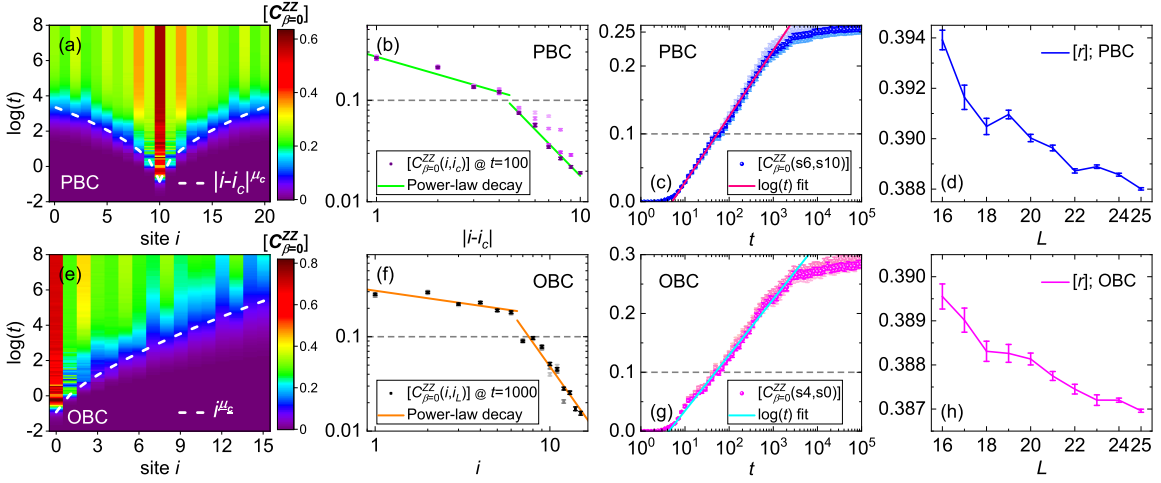}
\caption{\label{fig:fig1} cMBL's light cones via $[C^{\ZZ}_{\beta=0}]$, the ensemble-averaged OTOC. Here, the cMBL phase is ensured by selecting $\frac{g_x}{W_x}=0.9$ and $\frac{W_z}{W_x}=0$ \cite{ChenChen} whose level statistics \cite{Oganesyan} converges to Poisson under the finite-length scaling as shown by (d),(h). The first row [(a)-(d)] targets periodic boundary conditions (PBCs) with system size $L\!=\!21$, while the second row [(e)-(h)] implements the same calculation with $L\!=\!16$ under open boundary conditions (OBCs). In images (a),(e), white dashed lines delineate OTOC fronts as per a power-law fit between $\log(t)$ and operators' spacing. The obtained power-law exponent $\mu_c\!\approx\!0.539\ [\underline{\mu_c}\!\approx\!0.682]$ is less than $1$, so information spread inside cMBL is algebraically faster than in uMBL. The second column [(b),(f)] depicts spatial extent of the front, viz., horizontal cuts in (a),(e) along characteristic moments, from which a unique threshold $[C^{\ZZ}_{\beta=0}]\!\approx\!0.1$ (marked by grey dashed lines) is identified by noticing that OTOC above and below this value follows separately two individual power laws. The third column [(c),(g)], which stands for vertical cuts in (a),(e) on specified sites, illustrates the logarithmic temporal growth of OTOC in cMBL. Accompanied to the cusps of (b),(f), once OTOC exceeds the threshold, there arise the corresponding signatures of kinks in the time profiles of (c),(g). Light to solid colors in (b),(c) cover $L=15,17,19,21$ while in (f) [(g)] cover $L=12,14,16\ [L=16,18,20,22]$. Overall, finite-size effects are small for OTOCs in cMBL. Except for contour plots, all data points shown in this work are with error bars which represent standard deviations of the corresponding averaged quantities over more than $10^3$ quasirandom samples. All lines overlaid are obtained from the best fits of the selected data sets by weighted least-square method.}
\end{figure*}

{\em Numerical verification.}---Figure~\ref{fig:fig1} provides numerical evidence justifying the appropriateness of this new LR bound [Eq.~(\ref{LR_cMBL})] for cMBL. Specifically:
\begin{enumerate}
\item[(i)] The power-law fit between $\log(t)$ and operators' distance $|i-j|$ on the OTOC front [white dashed lines in Figs.~\ref{fig:fig1}(a),(e)] yields an exponent $\eta/\xi<1$ at a properly specified threshold, indicating information spreading in cMBL appears algebraically more efficient than that of uMBL where this exponent equals $1$.
\item[(ii)] The respective horizontal cuts across the light-cone images are presented by panels (b),(f) for two chosen instants, from which a threshold of the information front is identified at cusp where two separate power-law fits of OTOC data above and below this threshold intersect.
\item[(iii)] Figures~\ref{fig:fig1}(c) and (g) plot the characteristic logarithmic growth of OTOC, where kinks that appear when OTOC surpasses the threshold echo the cusps on the curves of the spatially power-law decay.
\end{enumerate}

\begin{figure}[b]
\centering
\includegraphics[width=1\linewidth]{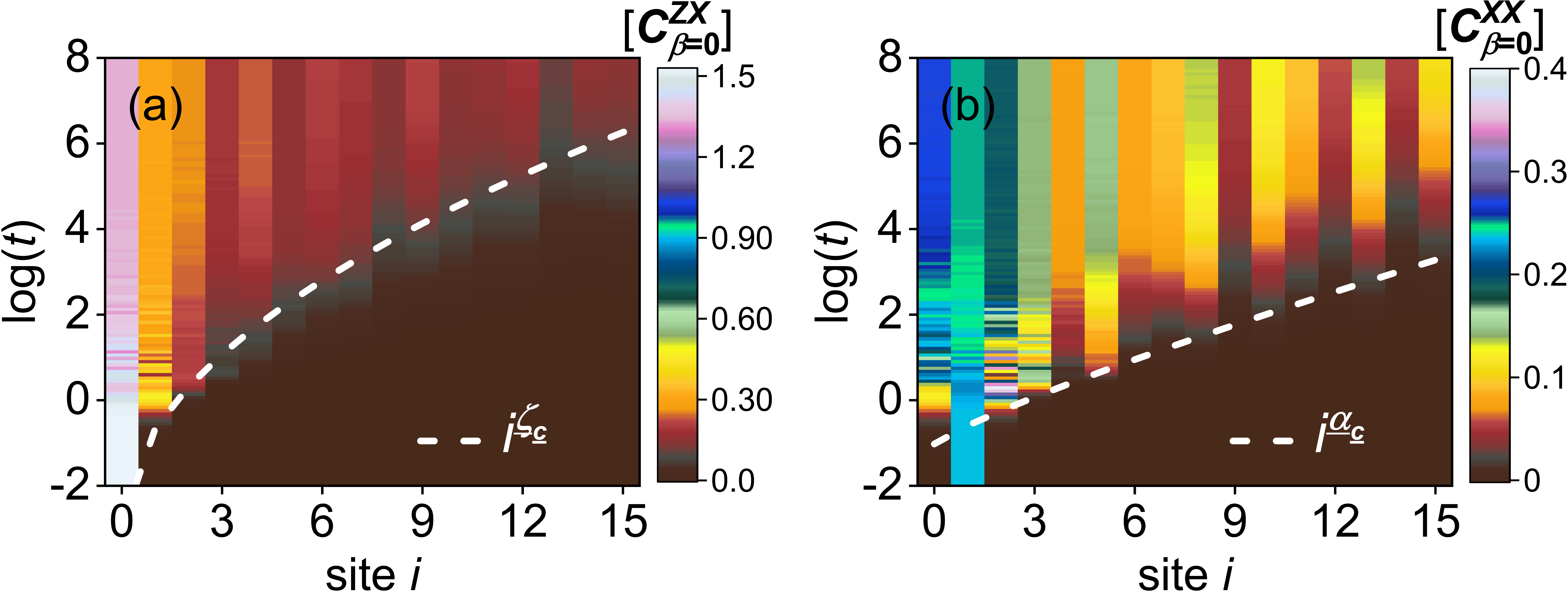}
\caption{\label{fig:fig2} cMBL's light cones via $[C^{\ZX}_{\beta=0}]$ and $[C^{\XX}_{\beta=0}]$ under OBCs. Hamiltonian parameters remain the same as in Fig.~\ref{fig:fig1}. Both power-law fitting exponents are less than $1$ with $\underline{\zeta_c}\approx0.516,\ \underline{\alpha_c}\approx0.849$.}
\end{figure}

Several comments are in order. First, the front shape, the spatial variation, and the temporal development of OTOC light cone are interrelated because determining any two of them settles the remaining one. For instance, in view of the reasonably good fits in panels [(a),(e)] and [(c),(g)], for consistency, the OTOC's spatial attenuation is anticipated to follow a power law. Second, the estimate of LR bound, in practice, mainly targets region close to light-cone front, therefore, considering finite-size fluctuations, it is possible that some elaborate choices of the threshold (like those indicated by kinks and cusps) are superior than other assignments. Third, the logarithmic temporal growth of OTOC constitutes one decisive result of current paper. As a direct implication, the establishment of such a logarithmic rise of OTOC reinforces the prediction on the double-log entanglement build-up in cMBL. Moreover, from either a numerical or experimental viewpoint, confirmation of such a single-log function is easier and more reliable than the detection of cMBL's double-log entanglement growth. Finally, in addition to $[C^{\ZZ}_{\beta=0}]$, the cMBL's $[C^{\ZX}_{\beta=0}]$ and $[C^{\XX}_{\beta=0}]$ components, as illustrated by Fig.~\ref{fig:fig2}, also show qualitatively similar dynamic behaviors.

The above numeric observations highlight the necessity of scrutinizing spatial and temporal profiles of OTOC contours on an equal footing because, for cMBL, although temporal growth of OTOC (entanglement entropy) slows to an unusual single-log (double-log) function of time, its spatial leakage of information is enhanced from an exponential suppression to a power-law decay. Take uMBL as a reference state, then for cMBL its temporal and spatial evolutions of OTOCs are tipped oppositely toward localization and thermalization; their conspiracy yields a smaller exponent $\eta/\xi<1$ for the light-cone front. In this regard, OTOCs paint a unified spacetime picture of information propagation and hint that cMBL may be a faster information scrambler than uMBL. Being consistent with our prior analysis on the embedded thermal-like core inside IOM of cMBL, this speculation does not necessarily contradict the phenomenology of full localization [see Figs.~\ref{fig:fig1}(d),(h)], because unlike the absolute disappearance of transports for conserved quantities such as energy and particle number, information spreading in general MBL systems continues in the thermodynamic limit and never ceases \cite{Abanin}. We hence resolve the skepticism about the compatibility between double-log entanglement growth and the emergent pronounced nonlocal correlations over confined length scales.

{\em Phenomenology derivation.}---Before embarking on LR bound for cMBL, it might be worthwhile to reexamine the applicability of the scenario involving the well-defined LIOMs \cite{HuseLIOM,Abanin,Serbyn,Ros}. Since model (\ref{constrham}) is local in projected Hilbert space, the na\"ive estimate of the LR bound produces a linear light cone typical for thermal states. Within uMBL, the LIOM scenario could come to the rescue, which posits that IOMs are not only commuting but also spatially quasilocal. Consequently, by rewriting a generic short-range disordered Hamiltonian in the IOM representation \cite{SuppMat},
\beq
H=\sum_{Z:\{j,N\}}h_{Z}=\sum_{Z:\{j,N\}}\sum_n\langle n|h_{Z}|n\rangle |n\rangle\langle n|,
\eeq
where $\{|n\rangle\}$ is the complete set of eigenstates and $\widetilde{h}_Z\!\coloneqq\!\sum_n\langle n|h_{Z}|n\rangle |n\rangle\langle n|$ is the IOM associated to the local term $h_Z$, one can readily derive a refined LR bound that engenders a logarithmic light cone,
\beq
\|[A(x,t),B(0,0)]\|_{\textrm{uMBL}}\lessapprox f'\frac{|t|}{e^{\varsigma |x|/\nu}}.
\label{LR_uMBL_2}
\eeq
Here the numerator $|t|$ results from the commutativity property $[\widetilde{h}_{Z'},\widetilde{h}_Z]\!=\!0$ and the asserted quasilocality of IOMs, viz.,
\beq
\sum_{Z\ni i,j}\|\widetilde{h}_Z\|\leqslant \lambda'_0\exp\!\left[-\frac{\varsigma}{\nu}\textrm{dist}(i,j)\right]\!,
\label{hz_exp}
\eeq
gives rise to the exponential in the denominator. The above derivation of (\ref{LR_uMBL_2}) was reported by \cite{Fan,HuangZhangChen,KimChandranAbanin}. For illustration, we reproduce their procedures of handling the key assumptions. Nevertheless, a parallel reasoning that mirrors the uMBL case does not work for cMBL. First, the numeric light cones from Figs.~\ref{fig:fig1}(a),(e) clearly violate the predictions of (\ref{LC_uMBL}) and (\ref{LR_uMBL_2}). Second, Figs.~\ref{fig:fig1}(b),(f) suggest that for cMBL the exponential fall-off in (\ref{hz_exp}) shall be replaced by a power law, then a straight application of the LIOM scenario yields an LR bound even looser than that of the thermal linear light cone, contradictory to our starting assumption on localization.

Phenomenologically, Ref.~\cite{ChenChen} reveals a nonnegligible thermal core enclosed by the IOM of cMBL, which naturally induces a length scale $\chi$ separating the differing short-range and long-range physics. Furthermore, this confined nonlocality, as demonstrated below, plays an active role in deriving the cMBL's LR bound, and thereby hints at the necessity to partially abandon the LIOM framework. The basic idea instead is to reformulate a Hastings-Koma (HK) series suitable for MBL by switching from the Heisenberg picture to the interaction picture where $\chi$ acquires a dynamical character.
 
To analytically address Eqs.~(\ref{LR_cMBL}) and (\ref{LC_cMBL}), we adopt the strategies of Refs.~\cite{HastingsKoma,FFG} and upgrade the original scheme from the few-body interactions in an ergodic system to the more general $k$-body interactions in the MBL setting.

A full derivation that leads to Eq.~(\ref{iter8}) below is detailed in \cite{SuppMat}. For conciseness, we summarize here the major involved rationale in the following $3$ successive steps. Concretely, we first divide the Hamiltonian (\ref{constrham}) into the short-range $[\textrm{diam}(Z^{sr})\leqslant\chi]$ and the long-range $[\textrm{diam}(Z^{lr})>\chi]$ parts,
\beq
H_{\textrm{qp}}=H^{sr}+H^{lr}=\sum_{Z^{sr}}\widetilde{h}_{Z^{sr}}+\sum_{Z^{lr}}\widetilde{h}_{Z^{lr}},
\eeq
according to which the interaction-picture operator reads $A_I(t)\!\coloneqq\!e^{iH^{sr}t}Ae^{-iH^{sr}t}$. As usual, the necessary connection between the Heisenberg and the interaction pictures is via the unitary scattering matrix, $A(t)\!=\!\mathcal{S}^{\dagger}(t)A_I(t)\mathcal{S}(t),\ \mathcal{S}(t)\!=\!e^{iH^{sr}t}e^{-iHt}$.

Next, the short-range contribution to the LR bound arising from the thermal core can be estimated by using the standard HK series,
\beq
\frac{\|[A_I(t),B]\|}{\|A\|\|B\|}\leqslant s|X|\exp\!\left[v|t|-\textrm{dist}(X,Y)/\chi(t)\right]\!,
\label{LRsr}
\eeq 
which indicates that the long-range operator $A_I(t)$ may be approximated by a sequence of intermediate operators whose supports are strictly finite-ranged,
\beq
\|A_I(\ell,t)-A_I(t)\|\leqslant s\|A\||X|\exp(-\ell),
\eeq
where $A_I(\ell,t)\coloneqq e^{iH^{sr}_\Lambda t}\{\int_{\widebar{\mathbb{B}}}d\mu(U) UAU^\dagger\}e^{-iH^{sr}_\Lambda t}$ \cite{Bravyi} and $\widebar{\mathbb{B}}$ denotes the complement to the ball $\mathbb{B}\coloneqq\{i\in\Lambda_s|\textrm{dist}(i,X)\leqslant R_\ell(t)\}$ whose radius $R_\ell(t)=R(t)+\ell\chi=\chi v|t|+\ell\chi,\ \ell=0,1,2,\ldots.$ Here, $X,Y$ represent the lattice sets holding the operators $A,B$, respectively.

Finally, the incorporation of the contributions from $H^{lr}$ entails the extension of the HK's scheme for the explicit inclusion of the overlapping conditions between the two disjoint intermediate operators, which bear the crucial $|t|$-dependence. Then, by invoking the discrete convolution amid the reduction of the augmented HK series, one obtains the following LR bound for the generic $k$-body Hamiltonian featuring the power-law decaying strengths,  
\begin{align}
\frac{\|[A(t),B]\|}{\|A\|\|B\|}&\leqslant s|X|(1+e)\left\{\frac{2e}{e-1}e^{vt-\textrm{dist}(X,Y)/\chi(t)} \right. \nonumber \\
&\left.+w^{-\eta}\cdot\frac{\exp\!\left\{g'\chi^{D-\eta} [R(t)]^D t\right\}}{\left[\textrm{dist}(X,Y)/R(t)\right]^\eta}\right\},
\label{iter8}
\end{align}
where $D$ is the spatial dimension, $|X|$ is the cardinality of set $X$, and $s,v,w,g'$ are positive coefficients. Up to prefactors, (\ref{iter8}) resembles the result of \cite{FFG}---the essential improvement relative to the HK's bound \cite{HastingsKoma} resides in the renormalization of the various contents under the influence of the emergent dynamical length scale $\chi$.

The derivation so far is somewhat quasi-rigorous except that it is based on certain fundamental assumptions that are of phenomenological nature. For example, both the form and the coefficients employed to describe the spatial distribution of the IOM's weights [like Eq.~(\ref{hz_exp})] are hypothesized because without exactly solving the model, they cannot be fully justified or expressed analytically in terms of the microscopic parameters. To make progress, usually one has to rely on the physical insights gained from finite-size numerics. The validity of the procedure can then be crosschecked by consistency. As will be illustrated, this reasoning carries over to the derivation of the key phenomenological result Eq.~(\ref{formula_chi}) below.

\begin{figure*}[tb]
\centering
\includegraphics[width=0.85\linewidth]{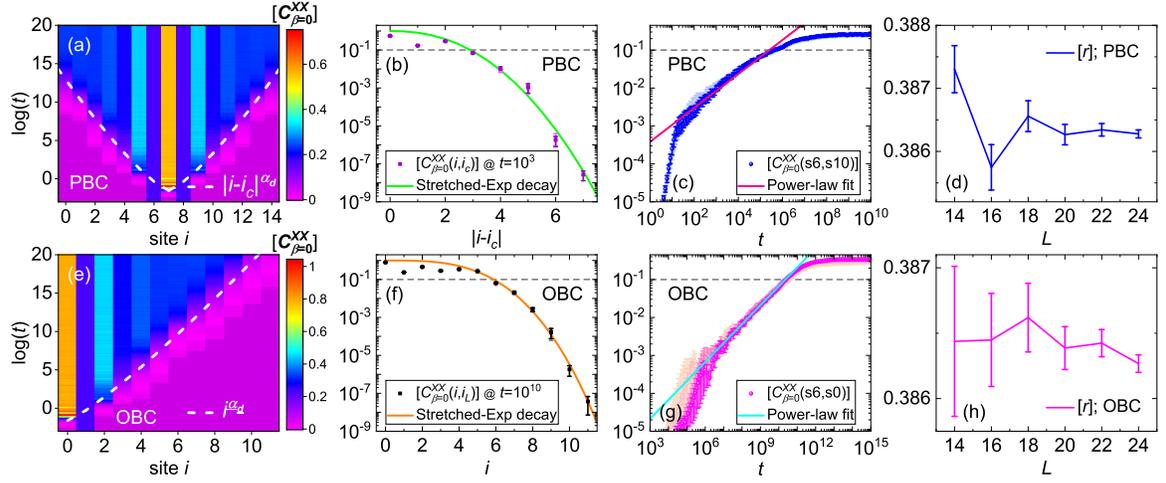}
\caption{\label{fig:fig3} dMBL's light cones via $[C^{\XX}_{\beta=0}]$. Here, the dMBL phase is stabilized by choosing $\frac{g_x}{W_x}=0.9$ and $\frac{W_z}{W_x}=8$ \cite{ChenChen} whose averaged level-spacing ratios are almost Poissonian, see (d),(h). [(a)-(d)] target PBCs with system size $L\!=\!15$; [(e)-(h)] execute the same calculation with $L\!=\!12$ utilizing OBCs. In (a),(e), the white dashed lines delimit the OTOC fronts (determined from the same threshold $[C^{\XX}_{\beta=0}]\!\approx\!0.1$) as per a power-law fit between $\log(t)$ and the operator separation. The obtained power-law exponent $\alpha_d\!\approx\!1.289\ (\underline{\alpha_d}\!\approx\!1.184)$ is greater than $1$, so information spread in dMBL is algebraically slower than that of uMBL. [(b),(f)] depict the spatial variation of the light cone along the horizontal cuts in (a),(e). The spatial attenuation of OTOC in dMBL is fitted by a stretched exponential. [(c),(g)] present the vertical cuts in (a),(e) at specified sites to exhibit the power-law temporal growth of OTOC inside dMBL. Light to solid colors of (b) [(f)] correspond to $L=11,13,15\ [L=8,10,12]$; while in (c) [(g)] they correspond to $L=15,17,19,21\ [L=8,12,16,20]$. Similarly, finite-size effects of OTOCs appear minor also in dMBL.}
\end{figure*}

In the case at hand, we do find such an ansatz solution for the phenomenological quantity $\chi$ as a function of time so that the simulated OTOC light-cone structures could be qualitatively comprehended to some extent. Indeed, specialized to cMBL, one finds that for $2D-1<\eta<2D$, by devising
\beq
\chi(t)=\rho(\ln t)^a t^b
\label{formula_chi}
\eeq
and substituting
\begin{align}
a&=-\frac{b}{D+1}=\frac{1}{2D-\eta}>1, \\
\rho&=\left[\frac{1}{v^D g'}\frac{\eta(\eta-D+1)}{2D-\eta}\right]^{\frac{1}{2D-\eta}},
\end{align}
the LR bound in (\ref{iter8}) simplifies to a desired form,
\begin{align}
\frac{\|[A(t),B]\|}{\|A\|\|B\|}&\leqslant s|X|(1+e)\!\left\{\frac{2e}{e-1}\exp\!\left[vt-\frac{\textrm{dist}(X,Y)}{\rho(\ln t)^at^{b}}\right] \right. \nonumber \\
&\left.+\left(\frac{v\rho}{w}\right)^{\eta}\cdot\frac{(\ln t)^{a\eta}}{\left[\textrm{dist}(X,Y)\right]^\eta}\right\}\!.
\label{iter9}
\end{align}
Specifically, Eq.~(\ref{iter9}) demonstrates that the OTOC of $A,B$ is bounded by the line $\textrm{dist}(X,Y)\propto(\ln t)^{a'}$ with $a'>a>1$, because for arbitrary threshold $\varepsilon$, there exists a critical moment $t_c<\infty$ such that whenever $\textrm{dist}(X,Y)\gtrapprox(\ln t)^{a'}$, $\|[A(t_c),B]\|<\varepsilon$. This rephrases asymptotically the relation (\ref{LC_cMBL}) of our numerical observations on cMBL. Meanwhile, the implication regarding the shrinkage of $\chi(t)$ is not only compatible with the general requisite for localization, but it also signals the potential stability of cMBL under thermodynamic limit because a confined thermal core has zero measure when system's space and time coordinates approach infinity.

To summarize, the overall rationale behind the above derivation is that by imposing the condition that the analytic formula (\ref{iter8}) explains the numeric OTOC light cones, an ansatz solution for the dynamical length scale can be obtained as (\ref{formula_chi}), which completes the main thread of the developed theory and renders the definition of cMBL self-consistent.

\vbox{} \noindent \textbf{Entropy bound}

\noindent This cMBL bound, together with the OTOC-R\'enyi-entropy theorem \cite{Fan,Hosur}, yields a bound for the rise of bipartite entropy on a finite open chain starting from randomized product states,
\beq
[S^{(2)}_{\textrm{R}}(t)]\lesssim-\ln(\vartheta-\varrho\ln t),
\eeq
where $\vartheta,\varrho$ are nonuniversal $t$-independent constants and $\vartheta>\varrho\ln t$ sets the saturation time scale, i.e., $t<t_{\textrm{sat}}$. Then, via elementary inequalities, it is straightforwardly proven that $\ln\ln(te^{\vartheta/\varrho})+\ln(\varrho/\vartheta^2)\leqslant-\ln(\vartheta-\varrho\ln t)$, implying that the observed double-log entanglement build-up in cMBL \cite{ChenChen} fulfills this entropy bound.

\vbox{} \noindent \textbf{LR bound in dMBL}

\begin{figure}[tb]
\centering
\includegraphics[width=1\linewidth]{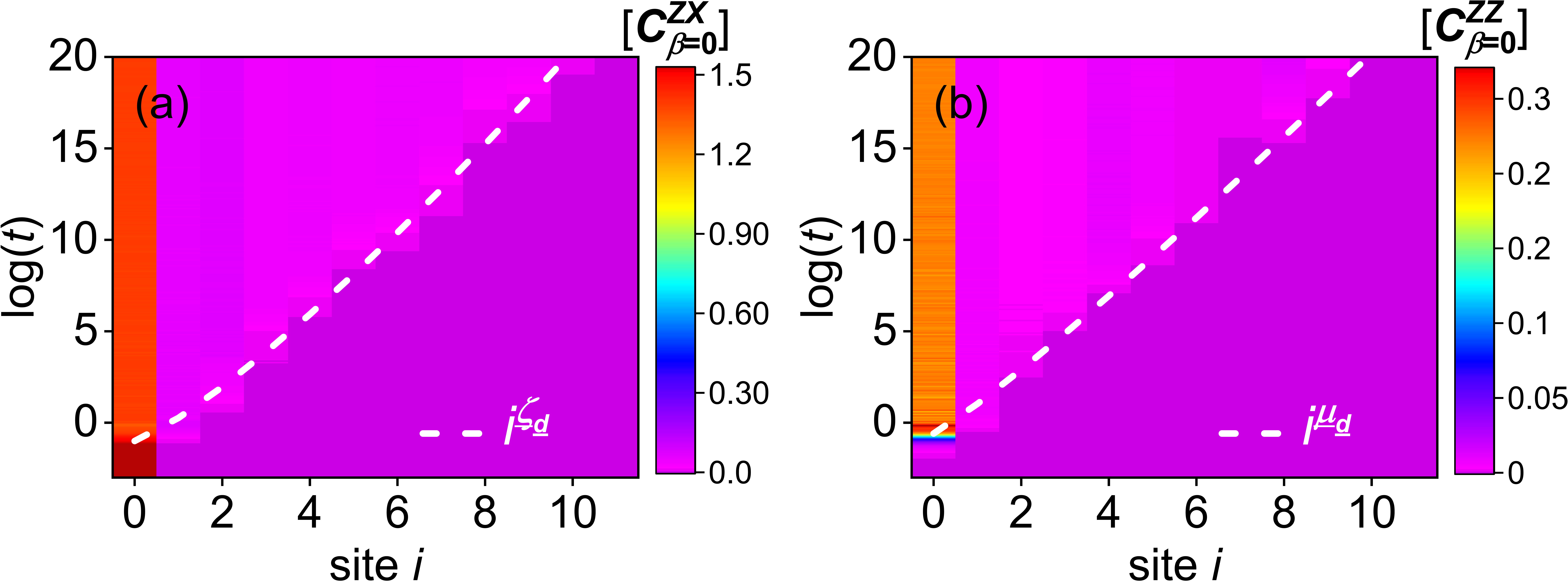}
\caption{\label{fig:fig4} dMBL's light cones via $[C^{\ZX}_{\beta=0}]$ and $[C^{\ZZ}_{\beta=0}]$ using OBCs. Hamiltonian parameters are the same as in Fig.~\ref{fig:fig3}. Two power-law fitting exponents are greater than $1$ with $\underline{\zeta_d}\approx1.228,\ \underline{\mu_d}\approx1.107$.}
\end{figure}

\noindent One central message of \cite{ChenChen} concerns the eigenstate transition between cMBL and diagonal MBL (dMBL) under the increase of $W_z$. Thrived upon that prediction, Fig.~\ref{fig:fig3} illustrates that the OTOC light cone of dMBL differs in fundamental aspects from that of cMBL.
\begin{enumerate}
\item[(i)] The power-law fit between $\log(t)$ and the operator distance along the OTOC front [white dashed lines in Figs.~\ref{fig:fig3}(a),(e)] delivers an exponent greater than $1$, suggesting the information transmission through dMBL is algebraically less efficient than that in uMBL.
\item[(ii)] The horizontal cuts of the light-cone images are displayed by (b),(f) for two representative moments, from which it is observed that to dMBL, the OTOC's spatial decay is delineated by a stretched exponential function, reflecting the fact that dMBL is a more robust localization phenomenon, consistent with the anticipation that constraints generally stymie certain intermediate channels of relaxation.
\item[(iii)] In line with uMBL, the temporal growth of OTOC in dMBL obeys a usual power-law function of time, as evidenced by Figs.~\ref{fig:fig3}(c),(g).
\end{enumerate}
For completeness, qualitatively analogous results on the OTOC contours for the dMBL's $[C^{\ZX}_{\beta=0}]$ and $[C^{\ZZ}_{\beta=0}]$ components are supplied by Fig.~\ref{fig:fig4}. Because IOMs of dMBL are dressed local $\widetilde{Z}_i$-operators \cite{Chen,ChenChen}, the saturated values of $[C^{\ZZ}_{\beta=0}]$ therein turn out to be significantly smaller than $[C^{\XX}_{\beta=0}]$ \cite{HuangZhangChen}. However, no such discrepancy arises in cMBL.

Conceptually, the dMBL phase is described by the LIOM phenomenology \cite{Chen,ChenChen}. Particularly, the above OTOC results can be comprehended to a large extent by introducing the following LR bound for dMBL,
\beq
\|[A(x,t),B(0,0)]\|_{\textrm{dMBL}}\lessapprox \widetilde{f} \exp[-\widetilde{\varsigma}|x|^{\widetilde{\kappa}}+\widetilde{\nu}\ln|t|].
\label{LB_dMBL}
\eeq
The derivation of (\ref{LB_dMBL}) mirrors that for (\ref{LR_uMBL_2}). Compared to the uMBL case, the essential difference lies in the functional change of the spatial weight from the canonical exponential decay in uMBL to the more severe stretched exponential fall-off in dMBL, as can be inferred from the condition that the exponent $\widetilde{\kappa}>1$. Therefore, after a minor modification of (\ref{hz_exp}) to $\sum_{Z\ni i,j}\|\widetilde{h}_Z\|\leqslant \widetilde{\lambda}'_0\exp[-\frac{\widetilde{\varsigma}}{\widetilde{\nu}}\textrm{dist}(i,j)^{\widetilde{\kappa}}]$, Eq.~(\ref{LB_dMBL}) follows immediately.

\vbox{} \noindent \textbf{Dynamics hierarchy}

\def\arraystretch{1.0}
\setlength{\tabcolsep}{9.5pt}
\begin{table}[b] 
\caption{Dynamics hierarchies of OTOC light-cone front, entanglement entropy, and quantum Fisher information spanning constrained, unconstrained, and diagonal MBL phases.} 
\label{table1} 
\centering
\begin{tabular}{ cccc }
 \\[-1.0em]
 \hline\hline
 \\[-0.9em]
 & LCF$_{[\textrm{OTOC}]}$ & $[S_\textrm{vN}]$ & $[$QFI$]$ \\ [0.1em] \cline{1-4} \\ [-0.9em]
 cMBL & $x\!\sim\![\log\left(t\right)]^{>1}$ & $\log\log\left(t\right)$ & $\log\log\log\left(t\right)$  \\ [0.1em] \cline{2-4} \\ [-0.9em]
 uMBL & $x\!\sim\![\log\left(t\right)]^{=1}$ & $\log\left(t\right)$ & $\log\log\left(t\right)$ \\[0.1em] \cline{2-4} \\ [-0.9em]
 dMBL & $x\!\sim\![\log\left(t\right)]^{<1}$ & $t^{\alpha}$ & $\log\left(t\right)$ \\[0.1em]
 \hline\hline
\end{tabular}
\end{table}

\noindent Table~\ref{table1} sums up dynamical features that single cMBL out as a distinctive state of matter relative to uMBL and dMBL, from which a phase-like hierarchy that encompasses cMBL, uMBL, and dMBL as well as an echelon relationship among OTOC, entanglement entropy, and quantum Fisher information (QFI) can be perceived. Here we define the von Neumann entanglement entropy upon tracing out degrees of freedom in half of the system, $S_{\textrm{vN}}(t)\coloneqq-\textrm{Tr}[\rho_R(t)\log_2\rho_R(t)]$, where $\rho_R$ is the reduced density matrix of the right half chain. Likewise, via initialization from the prepared N\'eel state $|\psi\rangle$, the QFI density acquires a simplified form analogous to the connected correlation function of staggered spin-imbalance operator $I\coloneqq\frac{1}{L}\sum_{i=1}^L(-1)^i\sigma^z_i$, i.e., $f_Q(t)\coloneqq4L[\langle\psi(t)|I^2|\psi(t)\rangle-\langle\psi(t)|I|\psi(t)\rangle^2]$. A detailed account on the time-evolving profiles of $[S_\textrm{vN}]$ and $[f_Q]$ was given in \cite{ChenChen}.

\vbox{} \noindent \textbf{DISCUSSION}

\noindent Experimentally, both the realization of the quasiperiodic version of the programmable Rydberg chain \cite{Bernien} by additionally imposing, for instance, the site-resolved potential offset \cite{SchreiberBloch,LukinGreiner} and the witness of OTOC and entanglement dynamics demand a high-fidelity local manipulation over the individual particles in analog quantum simulators by means of techniques such as spin echoes, optical tweezers, Rydberg states, nuclear spins, trapped ions, and quantum gas microscopes \cite{LiZhai,GarttnerRey,Meier,Landsman,Swingle,Wei}. Although tantalizingly challenging, direct detections of the characteristic double-log entanglement growth may still be attainable for cMBL using the developed protocols from Refs.~\cite{Kaufman,LukinGreiner,Brydges,Smith}. In parallel, it may be even more promising to probe the single-log build-up of the varied OTOCs as well as their power-law spatial decay to verify the cMBL phase and differentiate it from the dMBL and uMBL regimes.

To conclude, based on the experiment-inspired prototype minimal model, we compute the universal spacetime structures of OTOCs and LR bounds for the unconventional MBL states in constrained quantum spin systems. The OTOC in cMBL is characterized by a logarithmic temporal growth and a power-law spatial attenuation, whose light cone is captured by a new LR bound which we derive. In comparison, the dMBL phase is featured by a power-law rise of OTOC, which decreases in space as per a stretched exponential function of separation. These findings, along with the established uMBL phase, potentially point toward an intrinsic organization of the fully MBL states of matter. Could it be that a parent theory would foster a comprehensive classification of the hierarchical variety of the unconventional MBL quantum dynamics and beyond? The unified description and continued elucidation of these challenges may hold the prospect of enriching and advancing our current theoretical framework for localization.

\vbox{} \noindent \textbf{METHODS}

\noindent \textbf{Out-of-time-order commutators}

\noindent From the definition Eq.~(\ref{otoc_def}), it is easy to determine the following form of the out-of-time-order (OTO) commutator for the two Hermitian operators ${A}^\dagger(i,t)={A}(i,t)$ and ${B}^\dagger(j,t)={B}(j,t)$,
\begin{align}
C^{{A}{B}}_{\beta}(i,j;t,t')&=\frac{1}{2}\langle [{A}(i,t),{B}(j,t')]^\dagger [{A}(i,t),{B}(j,t')] \rangle_{\beta} \nonumber \\
&=\frac{1}{2}\langle {B}(j,t'){A}(i,t){A}(i,t){B}(j,t')\rangle_\beta \nonumber \\
&+\frac{1}{2}\langle {A}(i,t){B}(j,t'){B}(j,t'){A}(i,t)\rangle_\beta \nonumber \\
&-\langle {A}(i,t){B}(j,t'){A}(i,t){B}(j,t')\rangle_\beta,
\label{otoc_full}
\end{align}
where $\langle\mathcal{O}\rangle_\beta=\textrm{Tr}[e^{-\beta H_{\textrm{qp}}}\mathcal{O}]/\textrm{Tr}[e^{-\beta H_{\textrm{qp}}}]$ and in the last line the cyclic property of the trace is used. If, in addition, the Hermitian operators are unitary as well, i.e., ${A}^2(i,t)={B}^2(j,t)=\mathds{1}$, then
\beq
C^{{A}{B}}_{\beta}(i,j;t,t')=1-F^{{A}{B}}_{\beta}(i,j;t,t')
\eeq
with the OTO correlator defined by
\beq
F^{{A}{B}}_{\beta}(i,j;t,t')=\langle {A}(i,t){B}(j,t'){A}(i,t){B}(j,t')\rangle_\beta,
\eeq
which is apparently real in this case.

However, owing to the fact that within the projected Hilbert space, $\forall\ \mbox{site}\ i,\ \widetilde{X}_i^2\!\neq\!\mathds{1}$, the OTO commutators might not be exactly the same as the OTO correlators up to some constants. We are therefore chiefly relying on the general formula (\ref{otoc_full}) throughout the numerical evaluation of the various OTO commutators for the constrained Rydberg array.

\vbox{} \noindent \textbf{Exact diagonalization}

\noindent The intertwined complications of the many-body nonequilibrium problem that stem from superposed effects of constraint and randomness can be coped with by exact-diagonalization method for small $1$D finite systems. We resort to the standard full diagonalization algorithm to access the long-time behaviors of OTOCs, where quadruple precision is implemented to achieve the time evolution up to $t\approx 10^{20}$. Within full diagonalization, the infinite-time limit can be resolved by invoking the diagonal approximation.

\vbox{} \noindent \textbf{Hastings-Koma series}

\noindent To derive LR bound for cMBL [Eq.~(\ref{iter9})], we mainly borrow the general tactic of Hastings and Koma \cite{HastingsKoma} to exploit the nontrivial consequences derived from the locality structure of quantum many-body system when it is subject to constraint as well as irregularities \cite{SuppMat}. In particular, when deciphering OTOC constructions, rather than working with the bare lattice Hamiltonian, we reformulate a modified HK series using the integral-of-motion representation, which makes it possible to upgrade the original scheme from the two-body interactions in an ergodic situation to the $k$-body interactions suitable for cMBL. This improvement may find use in future applications.

\vbox{} \noindent \textbf{DATA AVAILABILITY}

\noindent The data set that supports the findings of the present study can be available from the corresponding authors via email upon reasonable request.

\vbox{} \noindent \textbf{ACKNOWLEDGEMENTS}

\noindent This work is supported by the National Key Research and Development Program of China (Grants Nos. 2017YFA0304204 and 2016YFA0300504), the National Natural Science Foundation of China Grant No. 11625416, and the Shanghai Municipal Government (Grants Nos. 19XD1400700 and 19JC1412702).

\vbox{} \noindent \textbf{AUTHOR CONTRIBUTIONS}

\noindent All authors contributed equally to this work.

\vbox{} \noindent \textbf{COMPETING INTERESTS}

\noindent The authors declare no competing financial or non-financial interests.

\bibliography{cMBL}

\begin{thebibliography}{40}%
\makeatletter
\providecommand \@ifxundefined [1]{%
 \@ifx{#1\undefined}
}%
\providecommand \@ifnum [1]{%
 \ifnum #1\expandafter \@firstoftwo
 \else \expandafter \@secondoftwo
 \fi
}%
\providecommand \@ifx [1]{%
 \ifx #1\expandafter \@firstoftwo
 \else \expandafter \@secondoftwo
 \fi
}%
\providecommand \natexlab [1]{#1}%
\providecommand \enquote  [1]{``#1''}%
\providecommand \bibnamefont  [1]{#1}%
\providecommand \bibfnamefont [1]{#1}%
\providecommand \citenamefont [1]{#1}%
\providecommand \href@noop [0]{\@secondoftwo}%
\providecommand \href [0]{\begingroup \@sanitize@url \@href}%
\providecommand \@href[1]{\@@startlink{#1}\@@href}%
\providecommand \@@href[1]{\endgroup#1\@@endlink}%
\providecommand \@sanitize@url [0]{\catcode `\\12\catcode `\$12\catcode
  `\&12\catcode `\#12\catcode `\^12\catcode `\_12\catcode `\%12\relax}%
\providecommand \@@startlink[1]{}%
\providecommand \@@endlink[0]{}%
\providecommand \url  [0]{\begingroup\@sanitize@url \@url }%
\providecommand \@url [1]{\endgroup\@href {#1}{\urlprefix }}%
\providecommand \urlprefix  [0]{URL }%
\providecommand \Eprint [0]{\href }%
\providecommand \doibase [0]{https://doi.org/}%
\providecommand \selectlanguage [0]{\@gobble}%
\providecommand \bibinfo  [0]{\@secondoftwo}%
\providecommand \bibfield  [0]{\@secondoftwo}%
\providecommand \translation [1]{[#1]}%
\providecommand \BibitemOpen [0]{}%
\providecommand \bibitemStop [0]{}%
\providecommand \bibitemNoStop [0]{.\EOS\space}%
\providecommand \EOS [0]{\spacefactor3000\relax}%
\providecommand \BibitemShut  [1]{\csname bibitem#1\endcsname}%
\let\auto@bib@innerbib\@empty
\bibitem [{\citenamefont {Anderson}(1958)}]{Anderson}%
  \BibitemOpen
  \bibfield  {author} {\bibinfo {author} {\bibfnamefont {P.~W.}\ \bibnamefont
  {Anderson}},\ }\bibfield  {title} {\bibinfo {title} {Absence of diffusion in
  certain random lattices},\ }\href {https://doi.org/10.1103/PhysRev.109.1492}
  {\bibfield  {journal} {\bibinfo  {journal} {Phys. Rev.}\ }\textbf {\bibinfo
  {volume} {109}},\ \bibinfo {pages} {1492} (\bibinfo {year}
  {1958})}\BibitemShut {NoStop}%
\bibitem [{\citenamefont {Basko}\ \emph {et~al.}(2006)\citenamefont {Basko},
  \citenamefont {Aleiner},\ and\ \citenamefont {Altshuler}}]{Basko}%
  \BibitemOpen
  \bibfield  {author} {\bibinfo {author} {\bibfnamefont {D.}~\bibnamefont
  {Basko}}, \bibinfo {author} {\bibfnamefont {I.}~\bibnamefont {Aleiner}},\
  and\ \bibinfo {author} {\bibfnamefont {B.}~\bibnamefont {Altshuler}},\
  }\bibfield  {title} {\bibinfo {title} {Metal-insulator transition in a weakly
  interacting many-electron system with localized single-particle states},\
  }\href {https://linkinghub.elsevier.com/retrieve/pii/S0003491605002630}
  {\bibfield  {journal} {\bibinfo  {journal} {Ann. Phys. (Amsterdam)}\ }\textbf
  {\bibinfo {volume} {321}},\ \bibinfo {pages} {1126} (\bibinfo {year}
  {2006})}\BibitemShut {NoStop}%
\bibitem [{\citenamefont {Gornyi}\ \emph {et~al.}(2005)\citenamefont {Gornyi},
  \citenamefont {Mirlin},\ and\ \citenamefont {Polyakov}}]{Gornyi}%
  \BibitemOpen
  \bibfield  {author} {\bibinfo {author} {\bibfnamefont {I.~V.}\ \bibnamefont
  {Gornyi}}, \bibinfo {author} {\bibfnamefont {A.~D.}\ \bibnamefont {Mirlin}},\
  and\ \bibinfo {author} {\bibfnamefont {D.~G.}\ \bibnamefont {Polyakov}},\
  }\bibfield  {title} {\bibinfo {title} {Interacting electrons in disordered
  wires: Anderson localization and low-{$T$} transport},\ }\href
  {https://doi.org/10.1103/PhysRevLett.95.206603} {\bibfield  {journal}
  {\bibinfo  {journal} {Phys. Rev. Lett.}\ }\textbf {\bibinfo {volume} {95}},\
  \bibinfo {pages} {206603} (\bibinfo {year} {2005})}\BibitemShut {NoStop}%
\bibitem [{\citenamefont {Oganesyan}\ and\ \citenamefont
  {Huse}(2007)}]{Oganesyan}%
  \BibitemOpen
  \bibfield  {author} {\bibinfo {author} {\bibfnamefont {V.}~\bibnamefont
  {Oganesyan}}\ and\ \bibinfo {author} {\bibfnamefont {D.~A.}\ \bibnamefont
  {Huse}},\ }\bibfield  {title} {\bibinfo {title} {Localization of interacting
  fermions at high temperature},\ }\href
  {https://doi.org/10.1103/PhysRevB.75.155111} {\bibfield  {journal} {\bibinfo
  {journal} {Phys. Rev. B}\ }\textbf {\bibinfo {volume} {75}},\ \bibinfo
  {pages} {155111} (\bibinfo {year} {2007})}\BibitemShut {NoStop}%
\bibitem [{\citenamefont {Abanin}\ \emph {et~al.}(2019)\citenamefont {Abanin},
  \citenamefont {Altman}, \citenamefont {Bloch},\ and\ \citenamefont
  {Serbyn}}]{Abanin}%
  \BibitemOpen
  \bibfield  {author} {\bibinfo {author} {\bibfnamefont {D.~A.}\ \bibnamefont
  {Abanin}}, \bibinfo {author} {\bibfnamefont {E.}~\bibnamefont {Altman}},
  \bibinfo {author} {\bibfnamefont {I.}~\bibnamefont {Bloch}},\ and\ \bibinfo
  {author} {\bibfnamefont {M.}~\bibnamefont {Serbyn}},\ }\bibfield  {title}
  {\bibinfo {title} {Colloquium: Many-body localization, thermalization, and
  entanglement},\ }\href {https://doi.org/10.1103/RevModPhys.91.021001}
  {\bibfield  {journal} {\bibinfo  {journal} {Rev. Mod. Phys.}\ }\textbf
  {\bibinfo {volume} {91}},\ \bibinfo {pages} {021001} (\bibinfo {year}
  {2019})}\BibitemShut {NoStop}%
\bibitem [{\citenamefont {De~Roeck}\ and\ \citenamefont
  {Huveneers}(2017)}]{DeRoeck}%
  \BibitemOpen
  \bibfield  {author} {\bibinfo {author} {\bibfnamefont {W.}~\bibnamefont
  {De~Roeck}}\ and\ \bibinfo {author} {\bibfnamefont {F.}~\bibnamefont
  {Huveneers}},\ }\bibfield  {title} {\bibinfo {title} {Stability and
  instability towards delocalization in many-body localization systems},\
  }\href {https://doi.org/10.1103/PhysRevB.95.155129} {\bibfield  {journal}
  {\bibinfo  {journal} {Phys. Rev. B}\ }\textbf {\bibinfo {volume} {95}},\
  \bibinfo {pages} {155129} (\bibinfo {year} {2017})}\BibitemShut {NoStop}%
\bibitem [{\citenamefont {\ifmmode~\check{S}\else \v{S}\fi{}untajs}\ \emph
  {et~al.}(2020)\citenamefont {\ifmmode~\check{S}\else \v{S}\fi{}untajs},
  \citenamefont {Bon\ifmmode~\check{c}\else \v{c}\fi{}a}, \citenamefont
  {Prosen},\ and\ \citenamefont {Vidmar}}]{SuntajsPRB}%
  \BibitemOpen
  \bibfield  {author} {\bibinfo {author} {\bibfnamefont {J.}~\bibnamefont
  {\ifmmode~\check{S}\else \v{S}\fi{}untajs}}, \bibinfo {author} {\bibfnamefont
  {J.}~\bibnamefont {Bon\ifmmode~\check{c}\else \v{c}\fi{}a}}, \bibinfo
  {author} {\bibfnamefont {T.}~\bibnamefont {Prosen}},\ and\ \bibinfo {author}
  {\bibfnamefont {L.}~\bibnamefont {Vidmar}},\ }\bibfield  {title} {\bibinfo
  {title} {Ergodicity breaking transition in finite disordered spin chains},\
  }\href {https://doi.org/10.1103/PhysRevB.102.064207} {\bibfield  {journal}
  {\bibinfo  {journal} {Phys. Rev. B}\ }\textbf {\bibinfo {volume} {102}},\
  \bibinfo {pages} {064207} (\bibinfo {year} {2020})}\BibitemShut {NoStop}%
\bibitem [{\citenamefont {Panda}\ \emph {et~al.}(2020)\citenamefont {Panda},
  \citenamefont {Scardicchio}, \citenamefont {Schulz}, \citenamefont {Taylor},\
  and\ \citenamefont {{\v{Z}}nidari{\v{c}}}}]{Panda_2020}%
  \BibitemOpen
  \bibfield  {author} {\bibinfo {author} {\bibfnamefont {R.~K.}\ \bibnamefont
  {Panda}}, \bibinfo {author} {\bibfnamefont {A.}~\bibnamefont {Scardicchio}},
  \bibinfo {author} {\bibfnamefont {M.}~\bibnamefont {Schulz}}, \bibinfo
  {author} {\bibfnamefont {S.~R.}\ \bibnamefont {Taylor}},\ and\ \bibinfo
  {author} {\bibfnamefont {M.}~\bibnamefont {{\v{Z}}nidari{\v{c}}}},\
  }\bibfield  {title} {\bibinfo {title} {Can we study the many-body
  localisation transition?},\ }\href
  {https://doi.org/10.1209/0295-5075/128/67003} {\bibfield  {journal} {\bibinfo
   {journal} {{EPL} (Europhysics Letters)}\ }\textbf {\bibinfo {volume}
  {128}},\ \bibinfo {pages} {67003} (\bibinfo {year} {2020})}\BibitemShut
  {NoStop}%
\bibitem [{\citenamefont {Sierant}\ \emph {et~al.}(2020)\citenamefont
  {Sierant}, \citenamefont {Delande},\ and\ \citenamefont
  {Zakrzewski}}]{SierantThoulessTime}%
  \BibitemOpen
  \bibfield  {author} {\bibinfo {author} {\bibfnamefont {P.}~\bibnamefont
  {Sierant}}, \bibinfo {author} {\bibfnamefont {D.}~\bibnamefont {Delande}},\
  and\ \bibinfo {author} {\bibfnamefont {J.}~\bibnamefont {Zakrzewski}},\
  }\bibfield  {title} {\bibinfo {title} {Thouless time analysis of {Anderson}
  and many-body localization transitions},\ }\href
  {https://doi.org/10.1103/PhysRevLett.124.186601} {\bibfield  {journal}
  {\bibinfo  {journal} {Phys. Rev. Lett.}\ }\textbf {\bibinfo {volume} {124}},\
  \bibinfo {pages} {186601} (\bibinfo {year} {2020})}\BibitemShut {NoStop}%
\bibitem [{\citenamefont {Abanin}\ \emph {et~al.}(2021)\citenamefont {Abanin},
  \citenamefont {Bardarson}, \citenamefont {{De Tomasi}}, \citenamefont
  {Gopalakrishnan}, \citenamefont {Khemani}, \citenamefont {Parameswaran},
  \citenamefont {Pollmann}, \citenamefont {Potter}, \citenamefont {Serbyn},\
  and\ \citenamefont {Vasseur}}]{Abanin2021}%
  \BibitemOpen
  \bibfield  {author} {\bibinfo {author} {\bibfnamefont {D.}~\bibnamefont
  {Abanin}}, \bibinfo {author} {\bibfnamefont {J.}~\bibnamefont {Bardarson}},
  \bibinfo {author} {\bibfnamefont {G.}~\bibnamefont {{De Tomasi}}}, \bibinfo
  {author} {\bibfnamefont {S.}~\bibnamefont {Gopalakrishnan}}, \bibinfo
  {author} {\bibfnamefont {V.}~\bibnamefont {Khemani}}, \bibinfo {author}
  {\bibfnamefont {S.}~\bibnamefont {Parameswaran}}, \bibinfo {author}
  {\bibfnamefont {F.}~\bibnamefont {Pollmann}}, \bibinfo {author}
  {\bibfnamefont {A.}~\bibnamefont {Potter}}, \bibinfo {author} {\bibfnamefont
  {M.}~\bibnamefont {Serbyn}},\ and\ \bibinfo {author} {\bibfnamefont
  {R.}~\bibnamefont {Vasseur}},\ }\bibfield  {title} {\bibinfo {title}
  {Distinguishing localization from chaos: Challenges in finite-size systems},\
  }\href {https://doi.org/https://doi.org/10.1016/j.aop.2021.168415} {\bibfield
   {journal} {\bibinfo  {journal} {Annals of Physics}\ }\textbf {\bibinfo
  {volume} {427}},\ \bibinfo {pages} {168415} (\bibinfo {year}
  {2021})}\BibitemShut {NoStop}%
\bibitem [{\citenamefont {\ifmmode \check{Z}\else
  \v{Z}\fi{}nidari\ifmmode~\check{c}\else \v{c}\fi{}}\ \emph
  {et~al.}(2008)\citenamefont {\ifmmode \check{Z}\else
  \v{Z}\fi{}nidari\ifmmode~\check{c}\else \v{c}\fi{}}, \citenamefont {Prosen},\
  and\ \citenamefont {Prelov\ifmmode~\check{s}\else \v{s}\fi{}ek}}]{Znidaric}%
  \BibitemOpen
  \bibfield  {author} {\bibinfo {author} {\bibfnamefont {M.}~\bibnamefont
  {\ifmmode \check{Z}\else \v{Z}\fi{}nidari\ifmmode~\check{c}\else
  \v{c}\fi{}}}, \bibinfo {author} {\bibfnamefont {T.}~\bibnamefont {Prosen}},\
  and\ \bibinfo {author} {\bibfnamefont {P.}~\bibnamefont
  {Prelov\ifmmode~\check{s}\else \v{s}\fi{}ek}},\ }\bibfield  {title} {\bibinfo
  {title} {Many-body localization in the {Heisenberg} {$XXZ$} magnet in a
  random field},\ }\href {https://doi.org/10.1103/PhysRevB.77.064426}
  {\bibfield  {journal} {\bibinfo  {journal} {Phys. Rev. B}\ }\textbf {\bibinfo
  {volume} {77}},\ \bibinfo {pages} {064426} (\bibinfo {year}
  {2008})}\BibitemShut {NoStop}%
\bibitem [{\citenamefont {Bardarson}\ \emph {et~al.}(2012)\citenamefont
  {Bardarson}, \citenamefont {Pollmann},\ and\ \citenamefont
  {Moore}}]{BardarsonPollmannMoore}%
  \BibitemOpen
  \bibfield  {author} {\bibinfo {author} {\bibfnamefont {J.~H.}\ \bibnamefont
  {Bardarson}}, \bibinfo {author} {\bibfnamefont {F.}~\bibnamefont
  {Pollmann}},\ and\ \bibinfo {author} {\bibfnamefont {J.~E.}\ \bibnamefont
  {Moore}},\ }\bibfield  {title} {\bibinfo {title} {Unbounded growth of
  entanglement in models of many-body localization},\ }\href
  {https://doi.org/10.1103/PhysRevLett.109.017202} {\bibfield  {journal}
  {\bibinfo  {journal} {Phys. Rev. Lett.}\ }\textbf {\bibinfo {volume} {109}},\
  \bibinfo {pages} {017202} (\bibinfo {year} {2012})}\BibitemShut {NoStop}%
\bibitem [{\citenamefont {Fan}\ \emph {et~al.}(2017)\citenamefont {Fan},
  \citenamefont {Zhang}, \citenamefont {Shen},\ and\ \citenamefont
  {Zhai}}]{Fan}%
  \BibitemOpen
  \bibfield  {author} {\bibinfo {author} {\bibfnamefont {R.}~\bibnamefont
  {Fan}}, \bibinfo {author} {\bibfnamefont {P.}~\bibnamefont {Zhang}}, \bibinfo
  {author} {\bibfnamefont {H.}~\bibnamefont {Shen}},\ and\ \bibinfo {author}
  {\bibfnamefont {H.}~\bibnamefont {Zhai}},\ }\bibfield  {title} {\bibinfo
  {title} {Out-of-time-order correlation for many-body localization},\ }\href
  {https://doi.org/https://doi.org/10.1016/j.scib.2017.04.011} {\bibfield
  {journal} {\bibinfo  {journal} {Science Bulletin}\ }\textbf {\bibinfo
  {volume} {62}},\ \bibinfo {pages} {707 } (\bibinfo {year}
  {2017})}\BibitemShut {NoStop}%
\bibitem [{\citenamefont {Huang}\ \emph {et~al.}(2016)\citenamefont {Huang},
  \citenamefont {Zhang},\ and\ \citenamefont {Chen}}]{HuangZhangChen}%
  \BibitemOpen
  \bibfield  {author} {\bibinfo {author} {\bibfnamefont {Y.}~\bibnamefont
  {Huang}}, \bibinfo {author} {\bibfnamefont {Y.-L.}\ \bibnamefont {Zhang}},\
  and\ \bibinfo {author} {\bibfnamefont {X.}~\bibnamefont {Chen}},\ }\bibfield
  {title} {\bibinfo {title} {Out-of-time-ordered correlators in many-body
  localized systems},\ }\href {https://doi.org/10.1002/andp.201600318}
  {\bibfield  {journal} {\bibinfo  {journal} {Annalen der Physik}\ }\textbf
  {\bibinfo {volume} {529}},\ \bibinfo {pages} {1600318} (\bibinfo {year}
  {2016})}\BibitemShut {NoStop}%
\bibitem [{\citenamefont {Huse}\ \emph {et~al.}(2014)\citenamefont {Huse},
  \citenamefont {Nandkishore},\ and\ \citenamefont {Oganesyan}}]{HuseLIOM}%
  \BibitemOpen
  \bibfield  {author} {\bibinfo {author} {\bibfnamefont {D.~A.}\ \bibnamefont
  {Huse}}, \bibinfo {author} {\bibfnamefont {R.}~\bibnamefont {Nandkishore}},\
  and\ \bibinfo {author} {\bibfnamefont {V.}~\bibnamefont {Oganesyan}},\
  }\bibfield  {title} {\bibinfo {title} {Phenomenology of fully
  many-body-localized systems},\ }\href
  {https://doi.org/10.1103/PhysRevB.90.174202} {\bibfield  {journal} {\bibinfo
  {journal} {Phys. Rev. B}\ }\textbf {\bibinfo {volume} {90}},\ \bibinfo
  {pages} {174202} (\bibinfo {year} {2014})}\BibitemShut {NoStop}%
\bibitem [{\citenamefont {Serbyn}\ \emph {et~al.}(2013)\citenamefont {Serbyn},
  \citenamefont {Papi\'c},\ and\ \citenamefont {Abanin}}]{Serbyn}%
  \BibitemOpen
  \bibfield  {author} {\bibinfo {author} {\bibfnamefont {M.}~\bibnamefont
  {Serbyn}}, \bibinfo {author} {\bibfnamefont {Z.}~\bibnamefont {Papi\'c}},\
  and\ \bibinfo {author} {\bibfnamefont {D.~A.}\ \bibnamefont {Abanin}},\
  }\bibfield  {title} {\bibinfo {title} {Local conservation laws and the
  structure of the many-body localized states},\ }\href
  {https://doi.org/10.1103/PhysRevLett.111.127201} {\bibfield  {journal}
  {\bibinfo  {journal} {Phys. Rev. Lett.}\ }\textbf {\bibinfo {volume} {111}},\
  \bibinfo {pages} {127201} (\bibinfo {year} {2013})}\BibitemShut {NoStop}%
\bibitem [{\citenamefont {Ros}\ \emph {et~al.}(2015)\citenamefont {Ros},
  \citenamefont {M\"uller},\ and\ \citenamefont {Scardicchio}}]{Ros}%
  \BibitemOpen
  \bibfield  {author} {\bibinfo {author} {\bibfnamefont {V.}~\bibnamefont
  {Ros}}, \bibinfo {author} {\bibfnamefont {M.}~\bibnamefont {M\"uller}},\ and\
  \bibinfo {author} {\bibfnamefont {A.}~\bibnamefont {Scardicchio}},\
  }\bibfield  {title} {\bibinfo {title} {Integrals of motion in the many-body
  localized phase},\ }\href
  {https://doi.org/https://doi.org/10.1016/j.nuclphysb.2014.12.014} {\bibfield
  {journal} {\bibinfo  {journal} {Nuclear Physics B}\ }\textbf {\bibinfo
  {volume} {891}},\ \bibinfo {pages} {420} (\bibinfo {year}
  {2015})}\BibitemShut {NoStop}%
\bibitem [{\citenamefont {Bernien}\ \emph {et~al.}(2017)\citenamefont
  {Bernien}, \citenamefont {Schwartz}, \citenamefont {Keesling}, \citenamefont
  {Levine}, \citenamefont {Omran}, \citenamefont {Pichler}, \citenamefont
  {Choi}, \citenamefont {Zibrov}, \citenamefont {Endres}, \citenamefont
  {Greiner}, \citenamefont {Vuleti\'c},\ and\ \citenamefont {Lukin}}]{Bernien}%
  \BibitemOpen
  \bibfield  {author} {\bibinfo {author} {\bibfnamefont {H.}~\bibnamefont
  {Bernien}}, \bibinfo {author} {\bibfnamefont {S.}~\bibnamefont {Schwartz}},
  \bibinfo {author} {\bibfnamefont {A.}~\bibnamefont {Keesling}}, \bibinfo
  {author} {\bibfnamefont {H.}~\bibnamefont {Levine}}, \bibinfo {author}
  {\bibfnamefont {A.}~\bibnamefont {Omran}}, \bibinfo {author} {\bibfnamefont
  {H.}~\bibnamefont {Pichler}}, \bibinfo {author} {\bibfnamefont
  {S.}~\bibnamefont {Choi}}, \bibinfo {author} {\bibfnamefont {A.~S.}\
  \bibnamefont {Zibrov}}, \bibinfo {author} {\bibfnamefont {M.}~\bibnamefont
  {Endres}}, \bibinfo {author} {\bibfnamefont {M.}~\bibnamefont {Greiner}},
  \bibinfo {author} {\bibfnamefont {V.}~\bibnamefont {Vuleti\'c}},\ and\
  \bibinfo {author} {\bibfnamefont {M.~D.}\ \bibnamefont {Lukin}},\ }\bibfield
  {title} {\bibinfo {title} {Probing many-body dynamics on a $51$-atom quantum
  simulator},\ }\href {https://www.nature.com/articles/nature24622} {\bibfield
  {journal} {\bibinfo  {journal} {Nature (London)}\ }\textbf {\bibinfo {volume}
  {551}},\ \bibinfo {pages} {579} (\bibinfo {year} {2017})}\BibitemShut
  {NoStop}%
\bibitem [{\citenamefont {Chen}\ \emph {et~al.}(2018)\citenamefont {Chen},
  \citenamefont {Burnell},\ and\ \citenamefont {Chandran}}]{Chen}%
  \BibitemOpen
  \bibfield  {author} {\bibinfo {author} {\bibfnamefont {C.}~\bibnamefont
  {Chen}}, \bibinfo {author} {\bibfnamefont {F.}~\bibnamefont {Burnell}},\ and\
  \bibinfo {author} {\bibfnamefont {A.}~\bibnamefont {Chandran}},\ }\bibfield
  {title} {\bibinfo {title} {How does a locally constrained quantum system
  localize?},\ }\href {https://doi.org/10.1103/PhysRevLett.121.085701}
  {\bibfield  {journal} {\bibinfo  {journal} {Phys. Rev. Lett.}\ }\textbf
  {\bibinfo {volume} {121}},\ \bibinfo {pages} {085701} (\bibinfo {year}
  {2018})}\BibitemShut {NoStop}%
\bibitem [{\citenamefont {Chen}\ \emph {et~al.}(2020)\citenamefont {Chen},
  \citenamefont {Chen},\ and\ \citenamefont {Wang}}]{ChenChen}%
  \BibitemOpen
  \bibfield  {author} {\bibinfo {author} {\bibfnamefont {C.}~\bibnamefont
  {Chen}}, \bibinfo {author} {\bibfnamefont {Y.}~\bibnamefont {Chen}},\ and\
  \bibinfo {author} {\bibfnamefont {X.}~\bibnamefont {Wang}},\ }\bibfield
  {title} {\bibinfo {title} {{Many-body localization in the
  infinite-interaction limit and the discontinuous eigenstate phase
  transition}},\ }\href {https://arxiv.org/abs/2011.09202} {\bibfield
  {journal} {\bibinfo  {journal} {arXiv:2011.09202}\ } (\bibinfo {year}
  {2020})}\BibitemShut {NoStop}%
\bibitem [{\citenamefont {Lieb}\ and\ \citenamefont
  {Robinson}(1972)}]{LiebRobinson}%
  \BibitemOpen
  \bibfield  {author} {\bibinfo {author} {\bibfnamefont {E.~H.}\ \bibnamefont
  {Lieb}}\ and\ \bibinfo {author} {\bibfnamefont {D.~W.}\ \bibnamefont
  {Robinson}},\ }\bibfield  {title} {\bibinfo {title} {{The finite group
  velocity of quantum spin systems}},\ }\href
  {https://doi.org/https://doi.org/10.1007/BF01645779} {\bibfield  {journal}
  {\bibinfo  {journal} {Commun. Math. Phys.}\ }\textbf {\bibinfo {volume}
  {28}},\ \bibinfo {pages} {251} (\bibinfo {year} {1972})}\BibitemShut
  {NoStop}%
\bibitem [{\citenamefont {Larkin}\ and\ \citenamefont
  {Ovchinnikov}(1969)}]{LarkinOvchinnikov}%
  \BibitemOpen
  \bibfield  {author} {\bibinfo {author} {\bibfnamefont {A.~I.}\ \bibnamefont
  {Larkin}}\ and\ \bibinfo {author} {\bibfnamefont {Y.~N.}\ \bibnamefont
  {Ovchinnikov}},\ }\bibfield  {title} {\bibinfo {title} {{Quasiclassical
  Method in the Theory of Superconductivity}},\ }\href
  {http://www.jetp.ac.ru/cgi-bin/e/index/e/28/6/p1200?a=list} {\bibfield
  {journal} {\bibinfo  {journal} {Sov. Phys. JETP}\ }\textbf {\bibinfo {volume}
  {28}},\ \bibinfo {pages} {120} (\bibinfo {year} {1969})}\BibitemShut
  {NoStop}%
\bibitem [{\citenamefont {Kitaev}(2015)}]{Kitaev}%
  \BibitemOpen
  \bibfield  {author} {\bibinfo {author} {\bibfnamefont {A.}~\bibnamefont
  {Kitaev}},\ }\bibfield  {title} {\bibinfo {title} {{A simple model of quantum
  holography}},\ }\href
  {https://online.kitp.ucsb.edu/online/entangled15/kitaev/} {\bibfield
  {journal} {\bibinfo  {journal} {Talks presented at the Kavli Institute for
  Theoretical Physics, University of California, Santa Barbara}\ } (\bibinfo
  {year} {7 April 2015 and 27 May 2015})}\BibitemShut {NoStop}%
\bibitem [{\citenamefont {Hosur}\ \emph {et~al.}(2016)\citenamefont {Hosur},
  \citenamefont {Qi}, \citenamefont {Roberts},\ and\ \citenamefont
  {Yoshida}}]{Hosur}%
  \BibitemOpen
  \bibfield  {author} {\bibinfo {author} {\bibfnamefont {P.}~\bibnamefont
  {Hosur}}, \bibinfo {author} {\bibfnamefont {X.-L.}\ \bibnamefont {Qi}},
  \bibinfo {author} {\bibfnamefont {D.~A.}\ \bibnamefont {Roberts}},\ and\
  \bibinfo {author} {\bibfnamefont {B.}~\bibnamefont {Yoshida}},\ }\bibfield
  {title} {\bibinfo {title} {{Chaos in quantum channels}},\ }\href
  {https://doi.org/10.1007/JHEP02(2016)004} {\bibfield  {journal} {\bibinfo
  {journal} {Journal of High Energy Phys.}\ }\textbf {\bibinfo {volume} {02}},\
  \bibinfo {pages} {004} (\bibinfo {year} {2016})}\BibitemShut {NoStop}%
\bibitem [{\citenamefont {Schreiber}\ \emph {et~al.}(2015)\citenamefont
  {Schreiber}, \citenamefont {Hodgman}, \citenamefont {Bordia}, \citenamefont
  {L\"uschen}, \citenamefont {Fischer}, \citenamefont {Vosk}, \citenamefont
  {Altman}, \citenamefont {Schneider},\ and\ \citenamefont
  {Bloch}}]{SchreiberBloch}%
  \BibitemOpen
  \bibfield  {author} {\bibinfo {author} {\bibfnamefont {M.}~\bibnamefont
  {Schreiber}}, \bibinfo {author} {\bibfnamefont {S.~S.}\ \bibnamefont
  {Hodgman}}, \bibinfo {author} {\bibfnamefont {P.}~\bibnamefont {Bordia}},
  \bibinfo {author} {\bibfnamefont {H.~P.}\ \bibnamefont {L\"uschen}}, \bibinfo
  {author} {\bibfnamefont {M.~H.}\ \bibnamefont {Fischer}}, \bibinfo {author}
  {\bibfnamefont {R.}~\bibnamefont {Vosk}}, \bibinfo {author} {\bibfnamefont
  {E.}~\bibnamefont {Altman}}, \bibinfo {author} {\bibfnamefont
  {U.}~\bibnamefont {Schneider}},\ and\ \bibinfo {author} {\bibfnamefont
  {I.}~\bibnamefont {Bloch}},\ }\bibfield  {title} {\bibinfo {title}
  {Observation of many-body localization of interacting fermions in a
  quasirandom optical lattice},\ }\href
  {https://doi.org/10.1126/science.aaa7432} {\bibfield  {journal} {\bibinfo
  {journal} {Science}\ }\textbf {\bibinfo {volume} {349}},\ \bibinfo {pages}
  {842} (\bibinfo {year} {2015})}\BibitemShut {NoStop}%
\bibitem [{\citenamefont {Lukin}\ \emph {et~al.}(2019)\citenamefont {Lukin},
  \citenamefont {Rispoli}, \citenamefont {Schittko}, \citenamefont {Tai},
  \citenamefont {Kaufman}, \citenamefont {Choi}, \citenamefont {Khemani},
  \citenamefont {L\'eonard},\ and\ \citenamefont {Greiner}}]{LukinGreiner}%
  \BibitemOpen
  \bibfield  {author} {\bibinfo {author} {\bibfnamefont {A.}~\bibnamefont
  {Lukin}}, \bibinfo {author} {\bibfnamefont {M.}~\bibnamefont {Rispoli}},
  \bibinfo {author} {\bibfnamefont {R.}~\bibnamefont {Schittko}}, \bibinfo
  {author} {\bibfnamefont {M.~E.}\ \bibnamefont {Tai}}, \bibinfo {author}
  {\bibfnamefont {A.~M.}\ \bibnamefont {Kaufman}}, \bibinfo {author}
  {\bibfnamefont {S.}~\bibnamefont {Choi}}, \bibinfo {author} {\bibfnamefont
  {V.}~\bibnamefont {Khemani}}, \bibinfo {author} {\bibfnamefont
  {J.}~\bibnamefont {L\'eonard}},\ and\ \bibinfo {author} {\bibfnamefont
  {M.}~\bibnamefont {Greiner}},\ }\bibfield  {title} {\bibinfo {title} {Probing
  entanglement in a many-body-localized system},\ }\href
  {https://science.sciencemag.org/content/364/6437/256.full} {\bibfield
  {journal} {\bibinfo  {journal} {Science}\ }\textbf {\bibinfo {volume}
  {364}},\ \bibinfo {pages} {256} (\bibinfo {year} {2019})}\BibitemShut
  {NoStop}%
\bibitem [{\citenamefont {Kim}\ \emph {et~al.}(2014)\citenamefont {Kim},
  \citenamefont {Chandran},\ and\ \citenamefont {Abanin}}]{KimChandranAbanin}%
  \BibitemOpen
  \bibfield  {author} {\bibinfo {author} {\bibfnamefont {I.~H.}\ \bibnamefont
  {Kim}}, \bibinfo {author} {\bibfnamefont {A.}~\bibnamefont {Chandran}},\ and\
  \bibinfo {author} {\bibfnamefont {D.~A.}\ \bibnamefont {Abanin}},\ }\bibfield
   {title} {\bibinfo {title} {Local integrals of motion and the logarithmic
  lightcone in many-body localized systems},\ }\href
  {https://arxiv.org/abs/1412.3073} {\bibfield  {journal} {\bibinfo  {journal}
  {arXiv:1412.3073}\ } (\bibinfo {year} {2014})}\BibitemShut {NoStop}%
\bibitem [{Sup()}]{SuppMat}%
  \BibitemOpen
  \href@noop {} {}\bibinfo {note} {See Supplementary information for the
  analytical derivation.}\BibitemShut {Stop}%
\bibitem [{\citenamefont {Hastings}\ and\ \citenamefont
  {Koma}(2006)}]{HastingsKoma}%
  \BibitemOpen
  \bibfield  {author} {\bibinfo {author} {\bibfnamefont {M.~B.}\ \bibnamefont
  {Hastings}}\ and\ \bibinfo {author} {\bibfnamefont {T.}~\bibnamefont
  {Koma}},\ }\bibfield  {title} {\bibinfo {title} {{Spectral Gap and
  Exponential Decay of Correlations}},\ }\href
  {https://doi.org/https://doi.org/10.1007/s00220-006-0030-4} {\bibfield
  {journal} {\bibinfo  {journal} {Commun. Math. Phys.}\ }\textbf {\bibinfo
  {volume} {265}},\ \bibinfo {pages} {781} (\bibinfo {year}
  {2006})}\BibitemShut {NoStop}%
\bibitem [{\citenamefont {Foss-Feig}\ \emph {et~al.}(2015)\citenamefont
  {Foss-Feig}, \citenamefont {Gong}, \citenamefont {Clark},\ and\ \citenamefont
  {Gorshkov}}]{FFG}%
  \BibitemOpen
  \bibfield  {author} {\bibinfo {author} {\bibfnamefont {M.}~\bibnamefont
  {Foss-Feig}}, \bibinfo {author} {\bibfnamefont {Z.-X.}\ \bibnamefont {Gong}},
  \bibinfo {author} {\bibfnamefont {C.~W.}\ \bibnamefont {Clark}},\ and\
  \bibinfo {author} {\bibfnamefont {A.~V.}\ \bibnamefont {Gorshkov}},\
  }\bibfield  {title} {\bibinfo {title} {{Nearly Linear Light Cones in
  Long-Range Interacting Quantum Systems}},\ }\href
  {https://doi.org/10.1103/PhysRevLett.114.157201} {\bibfield  {journal}
  {\bibinfo  {journal} {Phys. Rev. Lett.}\ }\textbf {\bibinfo {volume} {114}},\
  \bibinfo {pages} {157201} (\bibinfo {year} {2015})}\BibitemShut {NoStop}%
\bibitem [{\citenamefont {Bravyi}\ \emph {et~al.}(2006)\citenamefont {Bravyi},
  \citenamefont {Hastings},\ and\ \citenamefont {Verstraete}}]{Bravyi}%
  \BibitemOpen
  \bibfield  {author} {\bibinfo {author} {\bibfnamefont {S.}~\bibnamefont
  {Bravyi}}, \bibinfo {author} {\bibfnamefont {M.~B.}\ \bibnamefont
  {Hastings}},\ and\ \bibinfo {author} {\bibfnamefont {F.}~\bibnamefont
  {Verstraete}},\ }\bibfield  {title} {\bibinfo {title} {{Lieb-Robinson Bounds
  and the Generation of Correlations and Topological Quantum Order}},\ }\href
  {https://doi.org/10.1103/PhysRevLett.97.050401} {\bibfield  {journal}
  {\bibinfo  {journal} {Phys. Rev. Lett.}\ }\textbf {\bibinfo {volume} {97}},\
  \bibinfo {pages} {050401} (\bibinfo {year} {2006})}\BibitemShut {NoStop}%
\bibitem [{\citenamefont {Li}\ \emph {et~al.}(2017)\citenamefont {Li},
  \citenamefont {Fan}, \citenamefont {Wang}, \citenamefont {Ye}, \citenamefont
  {Zeng}, \citenamefont {Zhai}, \citenamefont {Peng},\ and\ \citenamefont
  {Du}}]{LiZhai}%
  \BibitemOpen
  \bibfield  {author} {\bibinfo {author} {\bibfnamefont {J.}~\bibnamefont
  {Li}}, \bibinfo {author} {\bibfnamefont {R.}~\bibnamefont {Fan}}, \bibinfo
  {author} {\bibfnamefont {H.}~\bibnamefont {Wang}}, \bibinfo {author}
  {\bibfnamefont {B.}~\bibnamefont {Ye}}, \bibinfo {author} {\bibfnamefont
  {B.}~\bibnamefont {Zeng}}, \bibinfo {author} {\bibfnamefont {H.}~\bibnamefont
  {Zhai}}, \bibinfo {author} {\bibfnamefont {X.}~\bibnamefont {Peng}},\ and\
  \bibinfo {author} {\bibfnamefont {J.}~\bibnamefont {Du}},\ }\bibfield
  {title} {\bibinfo {title} {Measuring out-of-time-order correlators on a
  nuclear magnetic resonance quantum simulator},\ }\href
  {https://doi.org/10.1103/PhysRevX.7.031011} {\bibfield  {journal} {\bibinfo
  {journal} {Phys. Rev. X}\ }\textbf {\bibinfo {volume} {7}},\ \bibinfo {pages}
  {031011} (\bibinfo {year} {2017})}\BibitemShut {NoStop}%
\bibitem [{\citenamefont {Garttner}\ \emph {et~al.}(2017)\citenamefont
  {Garttner}, \citenamefont {Bohnet}, \citenamefont {Safavi-Naini},
  \citenamefont {Wall}, \citenamefont {Bollinger},\ and\ \citenamefont
  {Rey}}]{GarttnerRey}%
  \BibitemOpen
  \bibfield  {author} {\bibinfo {author} {\bibfnamefont {M.}~\bibnamefont
  {Garttner}}, \bibinfo {author} {\bibfnamefont {J.~G.}\ \bibnamefont
  {Bohnet}}, \bibinfo {author} {\bibfnamefont {A.}~\bibnamefont
  {Safavi-Naini}}, \bibinfo {author} {\bibfnamefont {M.~L.}\ \bibnamefont
  {Wall}}, \bibinfo {author} {\bibfnamefont {J.~J.}\ \bibnamefont
  {Bollinger}},\ and\ \bibinfo {author} {\bibfnamefont {A.~M.}\ \bibnamefont
  {Rey}},\ }\bibfield  {title} {\bibinfo {title} {Measuring out-of-time-order
  correlations and multiple quantum spectra in a trapped-ion quantum magnet},\
  }\href {https://doi.org/10.1038/nphys4119} {\bibfield  {journal} {\bibinfo
  {journal} {Nat. Phys.}\ }\textbf {\bibinfo {volume} {13}},\ \bibinfo {pages}
  {781} (\bibinfo {year} {2017})}\BibitemShut {NoStop}%
\bibitem [{\citenamefont {Meier}\ \emph {et~al.}(2019)\citenamefont {Meier},
  \citenamefont {Ang'ong'a}, \citenamefont {An},\ and\ \citenamefont
  {Gadway}}]{Meier}%
  \BibitemOpen
  \bibfield  {author} {\bibinfo {author} {\bibfnamefont {E.~J.}\ \bibnamefont
  {Meier}}, \bibinfo {author} {\bibfnamefont {J.}~\bibnamefont {Ang'ong'a}},
  \bibinfo {author} {\bibfnamefont {F.~A.}\ \bibnamefont {An}},\ and\ \bibinfo
  {author} {\bibfnamefont {B.}~\bibnamefont {Gadway}},\ }\bibfield  {title}
  {\bibinfo {title} {Exploring quantum signatures of chaos on a floquet
  synthetic lattice},\ }\href {https://doi.org/10.1103/PhysRevA.100.013623}
  {\bibfield  {journal} {\bibinfo  {journal} {Phys. Rev. A}\ }\textbf {\bibinfo
  {volume} {100}},\ \bibinfo {pages} {013623} (\bibinfo {year}
  {2019})}\BibitemShut {NoStop}%
\bibitem [{\citenamefont {Landsman}\ \emph {et~al.}(2019)\citenamefont
  {Landsman}, \citenamefont {Figgatt}, \citenamefont {Schuster}, \citenamefont
  {Linke}, \citenamefont {Yoshida}, \citenamefont {Yao},\ and\ \citenamefont
  {Monroe}}]{Landsman}%
  \BibitemOpen
  \bibfield  {author} {\bibinfo {author} {\bibfnamefont {K.~A.}\ \bibnamefont
  {Landsman}}, \bibinfo {author} {\bibfnamefont {C.}~\bibnamefont {Figgatt}},
  \bibinfo {author} {\bibfnamefont {T.}~\bibnamefont {Schuster}}, \bibinfo
  {author} {\bibfnamefont {N.~M.}\ \bibnamefont {Linke}}, \bibinfo {author}
  {\bibfnamefont {B.}~\bibnamefont {Yoshida}}, \bibinfo {author} {\bibfnamefont
  {N.~Y.}\ \bibnamefont {Yao}},\ and\ \bibinfo {author} {\bibfnamefont
  {C.}~\bibnamefont {Monroe}},\ }\bibfield  {title} {\bibinfo {title} {Verified
  quantum information scrambling},\ }\href
  {https://www.nature.com/articles/s41586-019-0952-6} {\bibfield  {journal}
  {\bibinfo  {journal} {Nature (London)}\ }\textbf {\bibinfo {volume} {567}},\
  \bibinfo {pages} {61} (\bibinfo {year} {2019})}\BibitemShut {NoStop}%
\bibitem [{\citenamefont {Swingle}\ \emph {et~al.}(2016)\citenamefont
  {Swingle}, \citenamefont {Bentsen}, \citenamefont {Schleier-Smith},\ and\
  \citenamefont {Hayden}}]{Swingle}%
  \BibitemOpen
  \bibfield  {author} {\bibinfo {author} {\bibfnamefont {B.}~\bibnamefont
  {Swingle}}, \bibinfo {author} {\bibfnamefont {G.}~\bibnamefont {Bentsen}},
  \bibinfo {author} {\bibfnamefont {M.}~\bibnamefont {Schleier-Smith}},\ and\
  \bibinfo {author} {\bibfnamefont {P.}~\bibnamefont {Hayden}},\ }\bibfield
  {title} {\bibinfo {title} {Measuring the scrambling of quantum information},\
  }\href {https://doi.org/10.1103/PhysRevA.94.040302} {\bibfield  {journal}
  {\bibinfo  {journal} {Phys. Rev. A}\ }\textbf {\bibinfo {volume} {94}},\
  \bibinfo {pages} {040302} (\bibinfo {year} {2016})}\BibitemShut {NoStop}%
\bibitem [{\citenamefont {Wei}\ \emph {et~al.}(2018)\citenamefont {Wei},
  \citenamefont {Ramanathan},\ and\ \citenamefont {Cappellaro}}]{Wei}%
  \BibitemOpen
  \bibfield  {author} {\bibinfo {author} {\bibfnamefont {K.~X.}\ \bibnamefont
  {Wei}}, \bibinfo {author} {\bibfnamefont {C.}~\bibnamefont {Ramanathan}},\
  and\ \bibinfo {author} {\bibfnamefont {P.}~\bibnamefont {Cappellaro}},\
  }\bibfield  {title} {\bibinfo {title} {Exploring localization in nuclear spin
  chains},\ }\href {https://doi.org/10.1103/PhysRevLett.120.070501} {\bibfield
  {journal} {\bibinfo  {journal} {Phys. Rev. Lett.}\ }\textbf {\bibinfo
  {volume} {120}},\ \bibinfo {pages} {070501} (\bibinfo {year}
  {2018})}\BibitemShut {NoStop}%
\bibitem [{\citenamefont {Kaufman}\ \emph {et~al.}(2016)\citenamefont
  {Kaufman}, \citenamefont {Tai}, \citenamefont {Lukin}, \citenamefont
  {Rispoli}, \citenamefont {Schittko}, \citenamefont {Preiss},\ and\
  \citenamefont {Greiner}}]{Kaufman}%
  \BibitemOpen
  \bibfield  {author} {\bibinfo {author} {\bibfnamefont {A.~M.}\ \bibnamefont
  {Kaufman}}, \bibinfo {author} {\bibfnamefont {M.~E.}\ \bibnamefont {Tai}},
  \bibinfo {author} {\bibfnamefont {A.}~\bibnamefont {Lukin}}, \bibinfo
  {author} {\bibfnamefont {M.}~\bibnamefont {Rispoli}}, \bibinfo {author}
  {\bibfnamefont {R.}~\bibnamefont {Schittko}}, \bibinfo {author}
  {\bibfnamefont {P.~M.}\ \bibnamefont {Preiss}},\ and\ \bibinfo {author}
  {\bibfnamefont {M.}~\bibnamefont {Greiner}},\ }\bibfield  {title} {\bibinfo
  {title} {Quantum thermalization through entanglement in an isolated many-body
  system},\ }\href {https://doi.org/10.1126/science.aaf6725} {\bibfield
  {journal} {\bibinfo  {journal} {Science}\ }\textbf {\bibinfo {volume}
  {353}},\ \bibinfo {pages} {794} (\bibinfo {year} {2016})}\BibitemShut
  {NoStop}%
\bibitem [{\citenamefont {Brydges}\ \emph {et~al.}(2019)\citenamefont
  {Brydges}, \citenamefont {Elben}, \citenamefont {Jurcevic}, \citenamefont
  {Vermersch}, \citenamefont {Maier}, \citenamefont {Lanyon}, \citenamefont
  {Zoller}, \citenamefont {Blatt},\ and\ \citenamefont {Roos}}]{Brydges}%
  \BibitemOpen
  \bibfield  {author} {\bibinfo {author} {\bibfnamefont {T.}~\bibnamefont
  {Brydges}}, \bibinfo {author} {\bibfnamefont {A.}~\bibnamefont {Elben}},
  \bibinfo {author} {\bibfnamefont {P.}~\bibnamefont {Jurcevic}}, \bibinfo
  {author} {\bibfnamefont {B.}~\bibnamefont {Vermersch}}, \bibinfo {author}
  {\bibfnamefont {C.}~\bibnamefont {Maier}}, \bibinfo {author} {\bibfnamefont
  {B.~P.}\ \bibnamefont {Lanyon}}, \bibinfo {author} {\bibfnamefont
  {P.}~\bibnamefont {Zoller}}, \bibinfo {author} {\bibfnamefont
  {R.}~\bibnamefont {Blatt}},\ and\ \bibinfo {author} {\bibfnamefont {C.~F.}\
  \bibnamefont {Roos}},\ }\bibfield  {title} {\bibinfo {title} {{Probing
  R{\'e}nyi entanglement entropy via randomized measurements}},\ }\href
  {https://doi.org/10.1126/science.aau4963} {\bibfield  {journal} {\bibinfo
  {journal} {Science}\ }\textbf {\bibinfo {volume} {364}},\ \bibinfo {pages}
  {260} (\bibinfo {year} {2019})}\BibitemShut {NoStop}%
\bibitem [{\citenamefont {Smith}\ \emph {et~al.}(2016)\citenamefont {Smith},
  \citenamefont {Lee}, \citenamefont {Richerme}, \citenamefont {Neyenhuis},
  \citenamefont {Hess}, \citenamefont {Hauke}, \citenamefont {Heyl},
  \citenamefont {Huse},\ and\ \citenamefont {Monroe}}]{Smith}%
  \BibitemOpen
  \bibfield  {author} {\bibinfo {author} {\bibfnamefont {J.}~\bibnamefont
  {Smith}}, \bibinfo {author} {\bibfnamefont {A.}~\bibnamefont {Lee}}, \bibinfo
  {author} {\bibfnamefont {P.}~\bibnamefont {Richerme}}, \bibinfo {author}
  {\bibfnamefont {B.}~\bibnamefont {Neyenhuis}}, \bibinfo {author}
  {\bibfnamefont {P.~W.}\ \bibnamefont {Hess}}, \bibinfo {author}
  {\bibfnamefont {P.}~\bibnamefont {Hauke}}, \bibinfo {author} {\bibfnamefont
  {M.}~\bibnamefont {Heyl}}, \bibinfo {author} {\bibfnamefont {D.~A.}\
  \bibnamefont {Huse}},\ and\ \bibinfo {author} {\bibfnamefont
  {C.}~\bibnamefont {Monroe}},\ }\bibfield  {title} {\bibinfo {title}
  {Many-body localization in a quantum simulator with programmable random
  disorder},\ }\href {https://doi.org/10.1038/nphys3783} {\bibfield  {journal}
  {\bibinfo  {journal} {Nat. Phys.}\ }\textbf {\bibinfo {volume} {12}},\
  \bibinfo {pages} {907} (\bibinfo {year} {2016})}\BibitemShut {NoStop}%
\end{thebibliography}%


\begin{thebibliography}{4}%
\makeatletter
\providecommand \@ifxundefined [1]{%
 \@ifx{#1\undefined}
}%
\providecommand \@ifnum [1]{%
 \ifnum #1\expandafter \@firstoftwo
 \else \expandafter \@secondoftwo
 \fi
}%
\providecommand \@ifx [1]{%
 \ifx #1\expandafter \@firstoftwo
 \else \expandafter \@secondoftwo
 \fi
}%
\providecommand \natexlab [1]{#1}%
\providecommand \enquote  [1]{``#1''}%
\providecommand \bibnamefont  [1]{#1}%
\providecommand \bibfnamefont [1]{#1}%
\providecommand \citenamefont [1]{#1}%
\providecommand \href@noop [0]{\@secondoftwo}%
\providecommand \href [0]{\begingroup \@sanitize@url \@href}%
\providecommand \@href[1]{\@@startlink{#1}\@@href}%
\providecommand \@@href[1]{\endgroup#1\@@endlink}%
\providecommand \@sanitize@url [0]{\catcode `\\12\catcode `\$12\catcode
  `\&12\catcode `\#12\catcode `\^12\catcode `\_12\catcode `\%12\relax}%
\providecommand \@@startlink[1]{}%
\providecommand \@@endlink[0]{}%
\providecommand \url  [0]{\begingroup\@sanitize@url \@url }%
\providecommand \@url [1]{\endgroup\@href {#1}{\urlprefix }}%
\providecommand \urlprefix  [0]{URL }%
\providecommand \Eprint [0]{\href }%
\providecommand \doibase [0]{https://doi.org/}%
\providecommand \selectlanguage [0]{\@gobble}%
\providecommand \bibinfo  [0]{\@secondoftwo}%
\providecommand \bibfield  [0]{\@secondoftwo}%
\providecommand \translation [1]{[#1]}%
\providecommand \BibitemOpen [0]{}%
\providecommand \bibitemStop [0]{}%
\providecommand \bibitemNoStop [0]{.\EOS\space}%
\providecommand \EOS [0]{\spacefactor3000\relax}%
\providecommand \BibitemShut  [1]{\csname bibitem#1\endcsname}%
\let\auto@bib@innerbib\@empty
\bibitem [{\citenamefont {Bravyi}\ \emph {et~al.}(2006)\citenamefont {Bravyi},
  \citenamefont {Hastings},\ and\ \citenamefont {Verstraete}}]{Bravyi}%
  \BibitemOpen
  \bibfield  {author} {\bibinfo {author} {\bibfnamefont {S.}~\bibnamefont
  {Bravyi}}, \bibinfo {author} {\bibfnamefont {M.~B.}\ \bibnamefont
  {Hastings}},\ and\ \bibinfo {author} {\bibfnamefont {F.}~\bibnamefont
  {Verstraete}},\ }\bibfield  {title} {\bibinfo {title} {{Lieb-Robinson Bounds
  and the Generation of Correlations and Topological Quantum Order}},\ }\href
  {https://doi.org/10.1103/PhysRevLett.97.050401} {\bibfield  {journal}
  {\bibinfo  {journal} {Phys. Rev. Lett.}\ }\textbf {\bibinfo {volume} {97}},\
  \bibinfo {pages} {050401} (\bibinfo {year} {2006})}\BibitemShut {NoStop}%
\bibitem [{\citenamefont {Foss-Feig}\ \emph {et~al.}(2015)\citenamefont
  {Foss-Feig}, \citenamefont {Gong}, \citenamefont {Clark},\ and\ \citenamefont
  {Gorshkov}}]{FFG}%
  \BibitemOpen
  \bibfield  {author} {\bibinfo {author} {\bibfnamefont {M.}~\bibnamefont
  {Foss-Feig}}, \bibinfo {author} {\bibfnamefont {Z.-X.}\ \bibnamefont {Gong}},
  \bibinfo {author} {\bibfnamefont {C.~W.}\ \bibnamefont {Clark}},\ and\
  \bibinfo {author} {\bibfnamefont {A.~V.}\ \bibnamefont {Gorshkov}},\
  }\bibfield  {title} {\bibinfo {title} {{Nearly Linear Light Cones in
  Long-Range Interacting Quantum Systems}},\ }\href
  {https://doi.org/10.1103/PhysRevLett.114.157201} {\bibfield  {journal}
  {\bibinfo  {journal} {Phys. Rev. Lett.}\ }\textbf {\bibinfo {volume} {114}},\
  \bibinfo {pages} {157201} (\bibinfo {year} {2015})}\BibitemShut {NoStop}%
\bibitem [{\citenamefont {Nachtergaele}\ \emph {et~al.}(2006)\citenamefont
  {Nachtergaele}, \citenamefont {Ogata},\ and\ \citenamefont
  {Sims}}]{Nachtergaele}%
  \BibitemOpen
  \bibfield  {author} {\bibinfo {author} {\bibfnamefont {B.}~\bibnamefont
  {Nachtergaele}}, \bibinfo {author} {\bibfnamefont {Y.}~\bibnamefont
  {Ogata}},\ and\ \bibinfo {author} {\bibfnamefont {R.}~\bibnamefont {Sims}},\
  }\bibfield  {title} {\bibinfo {title} {{Propagation of Correlations in
  Quantum Lattice Systems}},\ }\href
  {https://doi.org/https://doi.org/10.1007/s10955-006-9143-6} {\bibfield
  {journal} {\bibinfo  {journal} {J. Stat. Phys.}\ }\textbf {\bibinfo {volume}
  {124}},\ \bibinfo {pages} {1} (\bibinfo {year} {2006})}\BibitemShut {NoStop}%
\bibitem [{\citenamefont {Hastings}\ and\ \citenamefont
  {Koma}(2006)}]{HastingsKoma}%
  \BibitemOpen
  \bibfield  {author} {\bibinfo {author} {\bibfnamefont {M.~B.}\ \bibnamefont
  {Hastings}}\ and\ \bibinfo {author} {\bibfnamefont {T.}~\bibnamefont
  {Koma}},\ }\bibfield  {title} {\bibinfo {title} {{Spectral Gap and
  Exponential Decay of Correlations}},\ }\href
  {https://doi.org/https://doi.org/10.1007/s00220-006-0030-4} {\bibfield
  {journal} {\bibinfo  {journal} {Commun. Math. Phys.}\ }\textbf {\bibinfo
  {volume} {265}},\ \bibinfo {pages} {781} (\bibinfo {year}
  {2006})}\BibitemShut {NoStop}%
\end{thebibliography}%

\end{document}


\title{Supplementary information for \lq\lq Lieb-Robinson bound for constrained many-body localization''}

\author{Chun~Chen}
\email[Corresponding author.\\]{chunchen@sjtu.edu.cn}
\affiliation{School of Physics and Astronomy, Key Laboratory of Artificial Structures and Quantum Control (Ministry of Education), Shenyang National Laboratory for Materials Science, Shanghai Jiao Tong University, Shanghai 200240, China}

\author{Xiaoqun~Wang}
\email[Corresponding author.\\]{xiaoqunwang@sjtu.edu.cn}
\affiliation{School of Physics and Astronomy, Key Laboratory of Artificial Structures and Quantum Control (Ministry of Education), Shenyang National Laboratory for Materials Science, Shanghai Jiao Tong University, Shanghai 200240, China}
\affiliation{Tsung-Dao Lee Institute, Shanghai Jiao Tong University, Shanghai 200240, China}
\affiliation{Beijing Computational Science Research Center, Beijing 100084, China}

\author{Yan~Chen}
\email[Corresponding author.\\]{yanchen99@fudan.edu.cn}
\affiliation{Department of Physics and State Key Laboratory of Surface Physics, Fudan University, Shanghai 200433, China}
\affiliation{Collaborative Innovation Center of Advanced Microstructures, Nanjing University, Nanjing 210093, China}

\date{\today}

\maketitle

\section{The Hamiltonian in the integral-of-motion representation}

It is an obvious fact that a simple, local, short-range, few-body Hamiltonian will typically acquire a very complicated, nonlocal, long-range, $k$-body form when rewritten in the representation of the integrals of motion. For example, assume the lattice system is described by the following Hamiltonian,
\beq
H=\sum_{\{j,N\}} h_{\{j,N\}},
\eeq
where $h_{\{j,N\}}$ is a locally interacting term centred on site $j$ and contains operators up to the $N$th nearest neighbors of $j$. Now introducing the unitary matrix $U$ that diagonalizes the matrix of the operator $H$ in a chosen basis, i.e.,
\beq
[UHU^\dagger]_{n',n}=\langle n'|H|n\rangle=E_n \langle n' | n \rangle=E_n\delta_{n',n}. 
\eeq
Then, in terms of the eigenenergies and the eigenstates, this relation simply means that
\beq
H=\sum_{\{j,N\}} \sum_n \langle n|h_{\{j,N\}}|n\rangle |n\rangle\langle n|=\sum_n E_n|n\rangle\langle n|,
\eeq
where $\left\{\sum_n \langle n|h_{\{j,N\}}|n\rangle |n\rangle\langle n|\right\}$ comprises the set of integrals of motion of the system, as they are not only mutually commuting but also commute with $H$. Normally, $\sum_n \langle n|h_{\{j,N\}}|n\rangle |n\rangle\langle n|$ is a highly nonlocal and $k$-body interacting term in the usual circumstances.

Specialize to the case of the unconstrained many-body-localized (uMBL) models, a widely-adopted theoretical framework to explain the breakdown of ergodicity there is based on the notion of the so-called quasilocal integrals of motion. Take, for instance, a general locally interacting spin-$1/2$ chain with randomness along $z$-direction, then according to this formalism, it is phenomenologically deduced that when the system is deeply inside the fully uMBL phase, the effective Hamiltonian matrix assumes
\beq
H_{\textrm{uMBL}}=\sum_i \widetilde{h}_i\tau^z_i+\sum_{i>j}J_{ij}\tau^z_i\tau^z_j+\sum_{i>j>k}J_{ijk}\tau^z_i\tau^z_j\tau^z_k+\ldots,
\eeq
where $\{[\tau^z_i]_{n',n}\coloneqq [U\sigma^z_iU^\dagger]_{n',n}\}$ comprises another set of quasilocal integrals of motion built from the physical spins $\{\sigma^z_i\}$ by the quasilocal unitary matrix $U$, which further implies that the various $k$-body interaction strengths $\{\widetilde{h}_i,J_{ij},J_{ijk},\ldots\}$ shall decay rapidly in space.

\section{Dividing the Hamiltonian into short-range and long-range parts}

After writing the local and finite-range Hamiltonian into a nonlocal and infinite-range form,
\beq
H=\sum_{\{j,N\}} h_{\{j,N\}}=\sum_{\{j,N\}} \widetilde{h}_{\{j,N\}},
\eeq
where the integrals of motion are defined by
\beq
\widetilde{h}_{\{j,N\}}\coloneqq \sum_n \langle n|h_{\{j,N\}}|n\rangle |n\rangle\langle n|,
\eeq 
one can proceed to separate $H$ into the short-range and the long-range pieces via imposing a dynamical length scale $\chi$. Namely,
\beq
H=\sum_{Z}\widetilde{h}_Z=\sum_{Z^{sr}}\widetilde{h}_{Z^{sr}}+\sum_{Z^{lr}}\widetilde{h}_{Z^{lr}}.
\label{Hsrlr}
\eeq
Here, any individual terms of $\{\widetilde{h}_Z\}$ that are supported exclusively on a lattice subset fulfilling $\textrm{diam}(Z^{sr})\leqslant\chi$ are grouped into the short-range part $\sum_{Z^{sr}}\widetilde{h}_{Z^{sr}}$. All the remaining terms are then classified as the long-range part $\sum_{Z^{lr}}\widetilde{h}_{Z^{lr}}$. Note that each $\widetilde{h}_Z$ might resemble $H$ in expression but they cannot be reduced to $h_Z$ because besides commuting with $H$, $\{\widetilde{h}_Z\}$ themselves also mutually commute.

\section{The interaction picture}

The main purpose of invoking Eq.~(\ref{Hsrlr}) is to transform the formulation of the Lieb-Robinson (LR) bound from the original Heisenberg picture to the interaction picture such that one is not only endowed with a variable length scale $\chi$ to tune but is also able to derive its meaningful form through the manipulation of the standard Hastings-Koma (HK) series. So, let's first introduce the apparatus of the interaction picture.

In order to be compatible with the OTOC results, we assume that the $k$-body Hamiltonian for the constrained MBL (cMBL) phase,
\beq
H_{\Lambda}=\sum_{Z\subset\Lambda_s}\widetilde{h}_Z,
\label{Ham}
\eeq
satisfies three conditions.
\begin{enumerate}
\item[(i)] The power-law decaying interactions, i.e.,
\beq
\sum_{Z\ni x,y}\|\widetilde{h}_Z\|\leqslant\frac{\lambda_0}{\left[1+\textrm{dist}\left(x,y\right)\right]^\eta},
\label{cond1}
\eeq
for any fixed site indices $x$ and $y$.
\item[(ii)] The reproducing condition,
\beq
\sum_{z\in\Lambda_s}\frac{1}{[1+\textrm{dist}(x,z)]^\eta}\frac{1}{[1+\textrm{dist}(z,y)]^\eta}\leqslant\frac{p_0}{[1+\textrm{dist}(x,y)]^\eta}.
\label{cond2}
\eeq
\item[(iii)] Related to the reproducing condition, it is assumed that
\beq
\sup_{\Lambda_s}\sup_x\sum_{y\in\Lambda_s}\frac{1}{\left[1+\textrm{dist}(x,y)\right]^\eta}=s'<\infty.
\label{cond3}
\eeq
\end{enumerate}
Here, positive constants $\lambda_0,\eta,p_0,s'$ are independent of the volume of $\Lambda_s$, the underlying lattice system of sites and bonds.

In accord with (\ref{Hsrlr}), the $k$-body interacting Hamiltonian (\ref{Ham}) is rewritten as follows,
\beq
H_{\Lambda}=H^{sr}_\Lambda+H^{lr}_\Lambda,
\eeq
where
\beq
H^{sr}_\Lambda=\sum_{\substack{Z^{sr}:Z^{sr}\subset\Lambda_s \\ \textrm{diam}(Z^{sr})\leqslant\chi}}\widetilde{h}_{Z^{sr}}\ \ \ \ \ \ \ \ \ \ \textrm{and}\ \ \ \ \ \ \ \ \ \ H^{lr}_\Lambda=\sum_{\substack{Z^{lr}:Z^{lr}\subset\Lambda_s \\ \textrm{diam}(Z^{lr})>\chi}}\widetilde{h}_{Z^{lr}}.
\eeq

Next, we recap the major formulas in different pictures. Throughout the derivation we always set $\hbar=1$. First, take $A$ as a generic time-independent Schr\"{o}dinger operator, then in the Heisenberg representation governed by $H_{\Lambda}$, it becomes
\beq
A_H(t)=e^{iH_{\Lambda}t}Ae^{-iH_{\Lambda}t}.
\eeq  
Instead, if choosing to evolve with just $H^{sr}_\Lambda$, one leads to the interaction-picture operator, 
\beq
A_I(t)=e^{iH^{sr}_{\Lambda}t}Ae^{-iH^{sr}_{\Lambda}t}.
\label{interactionops}
\eeq
It is easy to recognize the unitary scattering-matrix operator that connects the interaction picture to the Heisenberg picture,
\beq
A_H(t)=\mathcal{S}^\dagger(t)A_I(t)\mathcal{S}(t),\ \ \ \ \ \mathcal{S}(t)=e^{iH^{sr}_\Lambda t} e^{-iH_\Lambda t},
\eeq
whose form can be derived from its equation of motion,
\beq
\frac{d\mathcal{S}(t)}{dt}=H^{lr}_{\Lambda,I}(t)\mathcal{S}(t),\ \ \ \ \ H^{lr}_{\Lambda,I}(t)=e^{iH^{sr}_{\Lambda}t}H^{lr}_{\Lambda}e^{-iH^{sr}_{\Lambda}t},
\label{S_eom}
\eeq
as the following,
\beq
\mathcal{S}(t)=\sum^\infty_{n=0}\left(-\frac{i}{\hbar}\right)^n\frac{1}{n!}\int\limits^t_0dt_1\cdots\int\limits^t_0dt_n\ T\!\left[H^{lr}_{\Lambda,I}(t_1)H^{lr}_{\Lambda,I}(t_2)\cdots H^{lr}_{\Lambda,I}(t_n)\right].
\eeq

We start to analyze the short-range contribution to the LR bound from $H^{sr}_{\Lambda}$. Let $A,B$ be two observables supported on the compact sets, $X,Y\subset\Lambda_s$. The standard HK series gives
\begin{align}
\frac{\|[A_I(t),B]\|}{\|A\|}&\leqslant2\|B\|(2|t|)\sum_{\substack{Z^{sr}_1:Z^{sr}_1\cap X\neq\emptyset \\ Z^{sr}_1\cap Y\neq\emptyset}}\|\widetilde{h}_{Z^{sr}_1}\|+2\|B\|\frac{(2|t|)^2}{2!}\sum_{\substack{Z^{sr}_1:Z^{sr}_1\cap X\neq\emptyset}}\|\widetilde{h}_{Z^{sr}_1}\| \sum_{\substack{Z^{sr}_2:Z^{sr}_2\cap Z^{sr}_1\neq\emptyset \\ Z^{sr}_2\cap Y\neq\emptyset}}\|\widetilde{h}_{Z^{sr}_2}\| \nonumber \\
&+2\|B\|\frac{(2|t|)^3}{3!}\sum_{\substack{Z^{sr}_1:Z^{sr}_1\cap X\neq\emptyset}}\|\widetilde{h}_{Z^{sr}_1}\| \sum_{\substack{Z^{sr}_2:Z^{sr}_2\cap Z^{sr}_1\neq\emptyset}}\|\widetilde{h}_{Z^{sr}_2}\| \sum_{\substack{Z^{sr}_3:Z^{sr}_3\cap Z^{sr}_2\neq\emptyset \\ Z^{sr}_3\cap Y\neq\emptyset}}\|\widetilde{h}_{Z^{sr}_3}\|+\cdots.
\label{HKs}
\end{align}
By exploiting Eqs.~(\ref{cond1}), (\ref{cond2}), and (\ref{cond3}), one can simplify the terms of (\ref{HKs}) as follows,
\beq
\sum_{\substack{Z^{sr}_1:Z^{sr}_1\cap X\neq\emptyset \\ Z^{sr}_1\cap Y\neq\emptyset}}\|\widetilde{h}_{Z^{sr}_1}\|\leqslant\sum_{x\in X}\sum_{y\in Y}\sum_{Z^{sr}_1\ni x,y}\|\widetilde{h}_{Z^{sr}_1}\|\leqslant\sum_{x\in X}\sum_{y\in Y}\frac{\lambda_0}{\left[1+\textrm{dist}(x,y)\right]^\eta}\leqslant \sum_{x\in X}\lambda_0s'=|X|\lambda_0s'.
\eeq
For the second term,
\begin{align}
\sum_{\substack{Z^{sr}_1:Z^{sr}_1\cap X\neq\emptyset}}\|\widetilde{h}_{Z^{sr}_1}\| \sum_{\substack{Z^{sr}_2:Z^{sr}_2\cap Z^{sr}_1\neq\emptyset \\ Z^{sr}_2\cap Y\neq\emptyset}}\|\widetilde{h}_{Z^{sr}_2}\|&\leqslant\sum_{x\in X}\sum_{y\in Y}\sum_{z_{12}\in\Lambda_s}\sum_{Z^{sr}_1\ni x,z_{12}}\|\widetilde{h}_{Z^{sr}_1}\|\sum_{Z^{sr}_2\ni z_{12},y}\|\widetilde{h}_{Z^{sr}_2}\| \nonumber \\
&\leqslant\sum_{x\in X}\sum_{y\in Y}\sum_{z_{12}\in\Lambda_s}\frac{\lambda_0}{[1+\textrm{dist}(x,z_{12})]^\eta}\frac{\lambda_0}{[1+\textrm{dist}(z_{12},y)]^\eta} \nonumber \\
&\leqslant\sum_{x\in X}\sum_{y\in Y}\frac{\lambda^2_0p_0}{[1+\textrm{dist}(x,y)]^\eta}=|X|\lambda^2_0p_0s'.
\end{align}
Proceed analogously for the higher-order terms, one arrives at
\begin{align}
\frac{\|[A_I(t),B]\|}{\|A\|}&\leqslant2\|B\||X|p^{-1}_0s'\left(2|t|\lambda_0p_0+\frac{(2|t|\lambda_0p_0)^2}{2!}+\frac{(2|t|\lambda_0p_0)^3}{3!}+\cdots\right) \nonumber \\
&\leqslant2\|B\||X|p^{-1}_0s'\sum^\infty_{{\sf a}=\lceil r/\chi \rceil} \frac{(2|t|\lambda_0p_0)^{\sf a}}{{\sf a}!};\ \ \ \ \ \ r=\textrm{dist}(X,Y) \nonumber \\
&\leqslant2\|B\||X|p^{-1}_0s'\sum^\infty_{{\sf a}=\lceil r/\chi \rceil} \frac{(2|t|\lambda_0p_0)^{\sf a}}{{\sf a}!}e^{{\sf a}-\lceil r/\chi \rceil}\leqslant2\|B\||X|p^{-1}_0s'\sum^\infty_{{\sf a}=0} \frac{(2|t|e\lambda_0p_0)^{\sf a}}{{\sf a}!}e^{-\lceil r/\chi \rceil} \nonumber \\
&\leqslant2\|B\||X|p^{-1}_0s'\exp\!\left(2|t|e\lambda_0p_0-r/\chi\right),
\label{HKinteraction}
\end{align}
where in the second line, we use the observation that the range of $H^{sr}_\Lambda$ is within $\chi$ so that the HK sequence in (\ref{HKs}) needs at least $\lceil r/\chi\rceil$ iterations to bridge regions $X$ and $Y$, but in the final steps of (\ref{HKinteraction}) we still add back these contributions to get a closed expression.

Define the LR velocity for the short-range interactions,
\beq
v=2e\lambda_0p_0,
\eeq
Eq.~(\ref{HKinteraction}) is then repeated by
\beq
\frac{\|[A_I(t),B]\|}{\|A\|}\leqslant2\|B\||X|p^{-1}_0s'\exp\!\left(v|t|-r/\chi\right),
\label{LRsr}
\eeq
where picking up a threshold $\varepsilon_0$ gives rise to the definition of the radius $R(t)$ of the short-range linear light cone,
\beq
r=\chi v|t|-\chi\varepsilon_0\ \ \ \ \ \Longrightarrow\ \ \ \ \ R(t)=\chi v|t|.
\eeq

A common practice of using the result (\ref{LRsr}) is to approximate the long-range operator $A_I(t)$ by a sequence of intermediate operators whose supports, although expanding, are strictly finite-ranged. Following Refs.~\cite{Bravyi,FFG}, $\mathbb{B}[X,R_\ell(t)]$ denotes a ball of radius $R_\ell(t)=R(t)+\ell\chi,\ \ell=0,1,2,\ldots,$ centred on set $X$,
\beq
\mathbb{B}[X,R_\ell(t)]\coloneqq\left\{i\in\Lambda_s|\textrm{dist}(i,X)\leqslant R_\ell(t)\right\}.
\eeq
By an integration over the unitary group equipped with the Haar measure, the content of $A_I(t)$ confined within $\mathbb{B}[X,R_\ell(t)]$ can be formally isolated,
\beq
A_I(\ell,t)\coloneqq e^{iH^{sr}_\Lambda t}\left\{\ \int\limits_{\widebar{\mathbb{B}}[X,R_\ell(0)]}d\mu(U) UAU^\dagger\right\}e^{-iH^{sr}_\Lambda t},
\label{op_truncated}
\eeq    
where $\widebar{\mathbb{B}}[X,R_\ell(t)]$ is the complement of $\mathbb{B}[X,R_\ell(t)]$ with respect to $\Lambda_s$, i.e., operator $A_I(\ell,t)$ has no support outside the ball $\mathbb{B}[X,R(t)+\ell\chi(0)]$.

The usefulness of $\{A_I(\ell,t)\}$ resides in their resemblance to $A_I(t)$ in the sense that
\begin{align}
\left\|A_I(\ell,t)-A_I(t)\right\|&=\left\|\ e^{iH^{sr}_\Lambda t} \left\{\ \int\limits_{\widebar{\mathbb{B}}[X,R_\ell(0)]}d\mu(U) \left[UAU^\dagger-AUU^\dagger\right]\ \right\}e^{-iH^{sr}_\Lambda t}\right\| \nonumber \\
&\leqslant\ \int\limits_{\widebar{\mathbb{B}}[X,R_\ell(0)]}d\mu(U)\left\|\left[UA-AU\right]U^\dagger\right\|=\ \int\limits_{\widebar{\mathbb{B}}[X,R_\ell(0)]}d\mu(U)\left\|\left[A,U\right]\right\| \nonumber \\
&\leqslant 2\|A\|\|U\||X|p^{-1}_0s'\exp\!\left[-R_\ell(0)/\chi(0)\right]=2\|A\||X|p^{-1}_0s'\exp(-\ell),
\label{LRsr2}
\end{align}
where inequality (\ref{LRsr}) has been inserted and $\chi(0)$ should be understood as $\chi(t\rightarrow0^+)$. Due to the exponential reduction of the deviation as a function of $\ell$, when $\ell\rightarrow\infty$, $A_I(\infty,t)$ approaches $A_I(t)$ to a good approximation. Define next
\begin{align}
A^0_I(t)=A_I(0,t),\ \ \ \ \ A^{\ell>0}_I(t)=A_I(\ell,t)-A_I(\ell-1,t),
\end{align}
then clearly
\beq
A_I(t)=A_I(0,t)+\sum^\infty_{\ell=1}\left[A_I(\ell,t)-A_I(\ell-1,t)\right]=\sum^\infty_{\ell=0}A^\ell_I(t).
\label{decomp}
\eeq
Accordingly, relation (\ref{LRsr2}) implies that for $\ell>0$,
\begin{align}
\|A^\ell_I(t)\|&=\|A_I(\ell,t)-A_I(t)-[A_I(\ell-1,t)-A_I(t)]\| \nonumber \\
&\leqslant\|A_I(\ell,t)-A_I(t)\|+\|A_I(\ell-1,t)-A_I(t)\|\leqslant2\|A\||X|p^{-1}_0s'(1+e)\exp(-\ell).
\label{bound1}
\end{align}
Instead, for $\ell=0$,
\begin{align}
\|A^0_I(t)\|&\leqslant\|A_I(0,t)-A_I(t)\|+\|A_I(t)\|\leqslant2\|A\||X|p^{-1}_0s'+\|A\|.
\label{bound2}
\end{align}
Note that if $2|X|p^{-1}_0s'e\geqslant1$, then $\|A^0_I(t)\|\leqslant2\|A\||X|p^{-1}_0s'(1+e)$. In comparison, if $\|A\|\geqslant2|X|p^{-1}_0s'e\|A\|$, then the decay of the norm is even faster than the exponential for the first term.

Thus, as emphasized in \cite{FFG}, one shall in principle find a way to write $A_I(t)$ as a sum of a sequence of operators $A^\ell_I(t)$ with increasing (and finite) supports but exponentially decreasing bounds of the norm. This examination of the short-range interactions exemplifies the core content and the meaning of the employment of the interaction-picture formulation for the derivation of the LR bound.

\section{Incorporate long-range interactions in the interaction picture}

Our main target is the LR bound for the Heisenberg operators, therefore in terms of the scattering-matrix operator, it is easy to see that
\begin{align}
\left\|\left[A_H(t),B\right]\right\|&=\left\|\left[A_I(t),\mathcal{S}(t)B\mathcal{S}^\dagger(t)\right]\right\|\leqslant\sum^\infty_{\ell=0}\left\|\left[A^\ell_I(t),\mathcal{S}(t)B\mathcal{S}^\dagger(t)\right]\right\|.
\label{HeiCom}
\end{align}
Following \cite{FFG}, the overall strategy of tackling (\ref{HeiCom}) is to employ the interaction-picture equation of motion for $\mathcal{S}(t)$ to write a first-order differential equation for the relevant commutator defined below, and then the obtained result is converted into a HK-like integral equation that can be iterated to include the contributions from the long-range interactions in a progressive manner and thus produce the desired sequence of the full LR bound. Recall that $H^{lr}_\Lambda$ has been exclusively encapsulated in $\mathcal{S}(t)$.

Now, let's introduce the generalized two-time commutator,
\beq
G^\ell_r(t,\tau)\coloneqq\left[A^\ell_I(t),\mathcal{S}(\tau)B\mathcal{S}^\dagger(\tau)\right],\ \ \ \ \ r\coloneqq\textrm{dist}(X,Y).
\label{Gdef}
\eeq
Then, applying (\ref{S_eom}) yields,
\beq
\frac{dG^\ell_r(t,\tau)}{d\tau}=-i\left[A^\ell_I(t),\left[H^{lr}_{\Lambda,I}(\tau),\mathcal{S}(\tau)B\mathcal{S}^\dagger(\tau)\right]\right].
\label{G_eom}
\eeq
Implementing the same decomposition like that of (\ref{decomp}) for the long-range Hamiltonian,
\beq
H^{lr}_{\Lambda,I}(\tau)=\sum_{\substack{Z^{lr}}}\sum^\infty_{m=0}\widetilde{h}^m_{\ZlrI}(\tau),
\label{decomp_Hlr}
\eeq
where for each $Z^{lr}$, there arises a corresponding summation index $m_{Z^{lr}}$ whose subscript has been omitted in (\ref{decomp_Hlr}) for brevity. Then, it is easy to notice that the nested commutator is nonzero only if  there exist some $m$'s such that
\beq
\mathbb{B}[A^\ell_I(t)]\cap\mathbb{B}[\widetilde{h}^{m}_{\ZlrI}(\tau)]\neq\emptyset.
\label{condBall}
\eeq
Nonetheless, the mismatch between the arguments $t$ and $\tau$ renders the execution of this crucial condition very complicated in the ensuing HK iterations. More importantly, (\ref{condBall}) is the key resource where the explicit $t$-dependence can be extracted for modifying the LR bound. To this end, one would like to inquire whether there exists the kind of relations like the following,
\beq
\mathbb{B}[A^\ell_I(t)]\cap\mathbb{B}[\widetilde{h}^{m}_\ZlrI(\tau)]\neq\emptyset\ \ \ \Longrightarrow\ \ \ \mathbb{B}[A^\ell_I(t)]\cap\mathbb{B}[\widetilde{h}^{m}_\ZlrI(t)]\neq\emptyset\ ?
\eeq
As we now show, the answer is affirmative for the primary cases that interest us. The trick here is to utilize the familiar identity that for arbitrary operator $A$ and any unitary operator $U$,
\beq
\|A\|=\|UAU^\dagger\|,
\eeq
to shift the $t$-argument in the definition (\ref{interactionops}) of the interaction-picture operators. Concretely, for two arbitrary interaction-picture operators, one can prove that
\beq
\|[A_I(t),B_I(\tau)]\|=\|[A_I(t+\Delta t),B_I(\tau+\Delta t)]\|,
\label{t_shift}
\eeq
where the essential requirements are (1) $A_I(t)$ and $B_I(\tau)$ have to be defined by the same short-range Hamiltonian $H^{sr}_\Lambda$ and (2) Schr\"{o}dinger operators $A,B$ themselves are independent of time. Apparently, truncated operators generated by (\ref{op_truncated}) fulfill these two conditions.

The proof is divided into two parts: (I) the growing balls and (II) the shrinking balls.
\begin{enumerate}
\item[(I)] In the case of expanding balls, it is relatively easy to appreciate that based on
\beq
\mathbb{B}[A_I(\tau)]\cap\mathbb{B}[B_I(\tau')]\neq\emptyset,
\eeq
where $0\leqslant\tau',\tau\leqslant t$, one can safely pretend that this condition implies
\beq
\mathbb{B}[A_I(t)]\cap\mathbb{B}[B_I(t)]\neq\emptyset,
\label{set1}
\eeq
because the added contributions from extending the time domains are precisely compensated by the associated commutators (see below) which vanish identically in these enlarged regions of the balls. In other words, nothing extra has been included in essence. It is worth mentioning that this seemingly elementary extension from $\tau,\tau'$ to $t$ actually underpins the whole constructions of \cite{FFG}.  
\item[(II)] Because we concern the unconventional MBL, it is natural to anticipate that the balls or more precisely the dynamical length scale $\chi$ might be shrinking. For this circumstance, we proceed as follows: First,
\beq
[A_I(\tau),B_I(\tau')]\neq\emptyset\ \ \ \ \ \Longrightarrow\ \ \ \ \ \|[A_I(\tau),B_I(\tau')]\|\neq0.
\eeq
Then, shifting the arguments $\tau,\tau'$ by $t-\tau'$ as per (\ref{t_shift}),
\beq
\|[A_I(\tau),B_I(\tau')]\|=\|[A_I(\tau+t-\tau'),B_I(t)]\|\neq0,
\eeq
yields
\beq
\mathbb{B}[A_I(\tau)]\cap\mathbb{B}[B_I(\tau')]\neq\emptyset\ \ \ \Longrightarrow\ \ \ \mathbb{B}[A_I(\tau+t-\tau')]\cap\mathbb{B}[B_I(t)]\neq\emptyset.
\eeq
In the HK series (see below), typically
\beq
0\leqslant\tau'\leqslant\tau\leqslant t\ \ \ \ \ \Longrightarrow\ \ \ \ \ t\leqslant \tau+t-\tau'\leqslant \tau+t,
\eeq
therefore, for the case of shrinking balls,
\beq
\mathbb{B}[A_I(\tau)]\cap\mathbb{B}[B_I(\tau')]\neq\emptyset\ \ \ \Longrightarrow\ \ \ \mathbb{B}[A_I(t)]\cap\mathbb{B}[B_I(t)]\neq\emptyset
\label{set2}
\eeq
as desired. Apparently, (\ref{set2}) cannot be true for arbitrary $\tau$ and $\tau'$ with respect to $t$.
\end{enumerate}

Now let's turn to (\ref{G_eom}), whose expression can be transformed by the Jacobi identity, $[A,[B,C]]+[B,[C,A]]+[C,[A,B]]=0$, as
\begin{align}
\frac{dG^\ell_r(t,\tau)}{d\tau}=&-i\sum_{\substack{Z^{lr}}}\sum^\infty_{m=0}[\widetilde{h}^{m}_\ZlrI(\tau),[A^\ell_I(t),\mathcal{S}(\tau)B\mathcal{S}^\dagger(\tau)]]+i\sum_{\substack{Z^{lr}}}\sum^\infty_{m=0}[\mathcal{S}(\tau)B\mathcal{S}^\dagger(\tau),[A^\ell_I(t),\widetilde{h}^{m}_\ZlrI(\tau)]] \nonumber \\
=&-i\sum_{\substack{Z^{lr}}}\sum^\infty_{m=0}[\widetilde{h}^{m}_\ZlrI(\tau),[A^\ell_I(t),\mathcal{S}(\tau)B\mathcal{S}^\dagger(\tau)]]+i\sum_{\substack{Z^{lr}}}\sum^\infty_{m=0}\mathcal{D}_i(t;Z^{lr},m)[\mathcal{S}(\tau)B\mathcal{S}^\dagger(\tau),[A^\ell_I(t),\widetilde{h}^{m}_\ZlrI(\tau)]],
\label{Geom2}
\end{align}
where we introduce a step-like function \cite{FFG} as the knob to access the condition (\ref{condBall}),
\beq
\mathcal{D}_i(t;Z^{lr},m)=\left\{
\begin{array}{cl}
1 & \ \ \mbox{if}\ \ \ \mathbb{B}[A^\ell_I(t)]\cap\mathbb{B}[\widetilde{h}^{m}_\ZlrI(t)]\neq\emptyset, \\[.5em]
0 & \ \ \mbox{otherwise}.
\end{array}\right.
\label{dfactori}
\eeq
Implement again the Jacobi identity for the last term in (\ref{Geom2}) gives 
\begin{align}
\frac{dG^\ell_r(t,\tau)}{d\tau}=&-i[H^{lr}_{\Lambda,I}(\tau)-\widetilde{H}^{lr}_{\Lambda,I}(t,\tau),G^{lr}_r(t,\tau)]-i[A^{lr}_I(t),[\widetilde{H}^{lr}_{\Lambda,I}(t,\tau),\mathcal{S}(\tau)B\mathcal{S}^\dagger(\tau)]],
\label{Geom3}
\end{align}
where we define
\beq
\widetilde{H}^{lr}_{\Lambda,I}(t,\tau)\coloneqq\sum_{\substack{Z^{lr}}}\sum^\infty_{m=0}\mathcal{D}_i(t;Z^{lr},m)\widetilde{h}^{m}_\ZlrI(\tau).
\eeq

Then, a standard inequality in the theory of first-order differential equation \cite{Nachtergaele} applies to (\ref{Geom3}) and leads to the HK series in the interaction picture for iteration,
\begin{align}
\|[A^\ell_I(t),\mathcal{S}(t)B&\mathcal{S}^\dagger(t)]\|\leqslant\|[A^\ell_I(t),B]\|+\int\limits^t_0d\tau\|[A^\ell_I(t),[\widetilde{H}^{lr}_{\Lambda,I}(t,\tau),\mathcal{S}(\tau)B\mathcal{S}^\dagger(\tau)]]\| \nonumber \\
&\leqslant\|[A^\ell_I(t),B]\|+\sum_{Z^{lr},m}\mathcal{D}_i(t;Z^{lr},m)\int\limits^t_0d\tau\|[A^\ell_I(t),[\widetilde{h}^{m}_{\ZlrI}(\tau),\mathcal{S}(\tau)B\mathcal{S}^\dagger(\tau)]]\| \nonumber \\
&\leqslant\|[A^\ell_I(t),B]\|+2\|A^\ell_I(t)\|\sum_{Z^{lr},m}\mathcal{D}_i(t;Z^{lr},m)\int\limits^t_0d\tau\|[\widetilde{h}^{m}_{\ZlrI}(\tau),\mathcal{S}(\tau)B\mathcal{S}^\dagger(\tau)]\|,
\label{iter0}
\end{align}
where (\ref{Gdef}) has been recalled and $\sum_{Z^{lr},m}(\ldots)$ stands for $\sum_{\substack{Z^{lr}}}\sum^{\infty}_{m=0}(\ldots)$. Next, after recognizing the resemblance between $\|[\widetilde{h}^{m}_{\ZlrI}(\tau),\mathcal{S}(\tau)B\mathcal{S}^\dagger(\tau)]\|$ and $\|[A^\ell_I(t),\mathcal{S}(t)B\mathcal{S}^\dagger(t)]\|$, one might tend to start the iteration. A direct injection of the result (\ref{iter0}) leads to 
\begin{align}
\|[\widetilde{h}^{m}_{\ZlrI}(\tau),\mathcal{S}(\tau)&B\mathcal{S}^\dagger(\tau)]\|\leqslant\|[\widetilde{h}^{m}_{\ZlrI}(\tau),B]\| \nonumber \\
&+2\|\widetilde{h}^{m}_{\ZlrI}(\tau)\|\sum_{Z^{lr}_1,m_1}\mathcal{D}(\tau;Z^{lr},m;Z^{lr}_1,m_1)\int\limits^\tau_0d\tau_1\|[\widetilde{h}^{m_1}_{\ZlroI}(\tau_1),\mathcal{S}(\tau_1)B\mathcal{S}^\dagger(\tau_1)]\|,
\label{iter1}
\end{align}
where the $\tau$-dependence of $\mathcal{D}(\tau;Z^{lr},m;Z^{lr}_1,m_1)$ is inconvenient and insufficient for simplifying the expression of the iteration as more and more different $\tau_i$-dependence will be engendered. However, by going through the same derivations above, it is not hard to realize that $\mathcal{D}(\tau;Z^{lr},m;Z^{lr}_1,m_1)$ can be simply replaced by $\mathcal{D}(t;Z^{lr},m;Z^{lr}_1,m_1)$ in (\ref{iter1}).

For instance, within $0\leqslant\tau'\leqslant\tau\leqslant t$, define
\beq
\widetilde{G}^m_{\widetilde{r}}(\tau,\tau')\coloneqq[\widetilde{h}^m_{\ZlrI}(\tau),\mathcal{S}(\tau')B\mathcal{S}^\dagger(\tau')],\ \ \ \ \ \widetilde{r}\coloneqq\textrm{dist}(Z^{lr},Y).
\eeq
Parallel calculation using Eqs.~(\ref{set1}) and (\ref{set2}) yields
\beq
[\widetilde{h}^m_{\ZlrI}(\tau),H^{lr}_{\Lambda,I}(\tau')]=\sum_{Z^{lr}_1,m_1}\mathcal{D}(t;Z^{lr},m;Z^{lr}_1,m_1)[\widetilde{h}^m_{\ZlrI}(\tau),\widetilde{h}^{m_1}_{\ZlroI}(\tau')],
\eeq
where a general step function is given by
\beq
\mathcal{D}(t;Z^{lr},m;Z^{lr}_1,m_1)=\left\{
\begin{array}{cl}
1 & \ \ \mbox{if}\ \ \ \mathbb{B}[\widetilde{h}^{m}_\ZlrI(t)]\cap\mathbb{B}[\widetilde{h}^{m_1}_\ZlroI(t)]\neq\emptyset, \\[.5em]
0 & \ \ \mbox{otherwise}.
\end{array}\right.
\eeq
The resulting first-order differential equation of motion of $\widetilde{G}$ thus reads
\begin{align}
\frac{d\widetilde{G}^m_{\widetilde{r}}(\tau,\tau')}{d\tau'}=&-i[H^{lr}_{\Lambda,I}(\tau')-\widetilde{H}^{Z^{lr}\!\!,m}_{\Lambda,I}(t,\tau'),\widetilde{G}^{m}_{\widetilde{r}}(\tau,\tau')] \nonumber \\
&-i\sum_{Z^{lr}_1,m_1}\mathcal{D}(t;Z^{lr},m;Z^{lr}_1,m_1)[\widetilde{h}^{m}_\ZlrI(\tau),[\widetilde{h}^{m_1}_{\ZlroI}(\tau'),\mathcal{S}(\tau')B\mathcal{S}^\dagger(\tau')]],
\end{align}
where we define
\beq
\widetilde{H}^{Z^{lr}\!\!,m}_{\Lambda,I}(t,\tau')\coloneqq\sum_{\substack{Z^{lr}_1}}\sum^\infty_{m_1=0}\mathcal{D}(t;Z^{lr},m;Z^{lr}_1,m_1)\widetilde{h}^{m_1}_\ZlroI(\tau').
\eeq
Applying again the theorem in \cite{Nachtergaele} yields
\begin{align}
\|\widetilde{G}^m_{\widetilde{r}}(\tau,\tau)\|&\leqslant\|\widetilde{G}^m_{\widetilde{r}}(\tau,0)\|+\int\limits^\tau_0d\tau'\sum_{Z^{lr}_1,m_1}\mathcal{D}(t;Z^{lr},m;Z^{lr}_1,m_1)\|[\widetilde{h}^{m}_\ZlrI(\tau),[\widetilde{h}^{m_1}_{\ZlroI}(\tau'),\mathcal{S}(\tau')B\mathcal{S}^\dagger(\tau')]]\| \nonumber \\
&\leqslant\|\widetilde{G}^m_{\widetilde{r}}(\tau,0)\|+\sum_{Z^{lr}_1,m_1}\mathcal{D}(t;Z^{lr},m;Z^{lr}_1,m_1)2\|\widetilde{h}^{m}_\ZlrI\|\int\limits^\tau_0d\tau'\|[\widetilde{h}^{m_1}_{\ZlroI}(\tau'),\mathcal{S}(\tau')B\mathcal{S}^\dagger(\tau')]\|,
\label{iter2}
\end{align}
where $\|\widetilde{h}^{m}_\ZlrI(\tau)\|=\|\widetilde{h}^{m}_\ZlrI\|$ is time independent as per (\ref{op_truncated}), or otherwise it can be replaced by the corresponding time-independent bound as per (\ref{bound1}) and (\ref{bound2}). Clearly, (\ref{iter2}) comprises the desired relation for initiating all the remaining iterations.

Armed with these preparations, one can now derive the generalized HK series in the interaction picture by repeatedly inserting (\ref{iter2}) into (\ref{iter0}). This iterative procedure yields the following,
\begin{align}
\|[A^\ell_I&(t),\mathcal{S}(t)B\mathcal{S}^\dagger(t)]\|\leqslant\|[A^\ell_I(t),B]\|+2\|A^\ell_I(t)\|\sum_{Z^{lr}_0,m_0}\mathcal{D}_i(t;Z^{lr}_0,m_0)2\|\widetilde{h}^{m_0}_{\ZlrzI}\|\mathcal{D}_f(t;Z^{lr}_0,m_0)\|B\|\int\limits^t_0d\tau_0 \nonumber \\
&+2\|A^\ell_I(t)\|\sum_{Z^{lr}_0,m_0}\mathcal{D}_i(t;Z^{lr}_0,m_0)2\|\widetilde{h}^{m_0}_{\ZlrzI}\|\sum_{Z^{lr}_1,m_1}\mathcal{D}(t;Z^{lr}_0,m_0;Z^{lr}_1,m_1)2\|\widetilde{h}^{m_1}_{\ZlroI}\| \nonumber \\
&\times\mathcal{D}_f(t;Z^{lr}_1,m_1)\|B\|\int\limits^t_0d\tau_0\int\limits^{\tau_0}_0 d\tau_1+2\|A^\ell_I(t)\|\sum_{Z^{lr}_0,m_0}\mathcal{D}_i(t;Z^{lr}_0,m_0)2\|\widetilde{h}^{m_0}_{\ZlrzI}\|\sum_{Z^{lr}_1,m_1}\mathcal{D}(t;Z^{lr}_0,m_0;Z^{lr}_1,m_1)2\|\widetilde{h}^{m_1}_{\ZlroI}\| \nonumber \\
&\times\sum_{Z^{lr}_2,m_2}\mathcal{D}(t;Z^{lr}_1,m_1;Z^{lr}_2,m_2)2\|\widetilde{h}^{m_2}_{\ZlrtI}\|\mathcal{D}_f(t;Z^{lr}_2,m_2)\|B\|\int\limits^t_0d\tau_0\int\limits^{\tau_0}_0 d\tau_1\int\limits^{\tau_1}_0 d\tau_2+\cdots,
\label{iter3}
\end{align}
where similar to (\ref{dfactori}), symbolically
\beq
\mathcal{D}_f(t;Z^{lr},m)=\left\{
\begin{array}{cl}
1 & \ \ \mbox{if}\ \ \ \mathbb{B}[\widetilde{h}^{m}_\ZlrI(t)]\cap\mathbb{B}[B_I(t)]\neq\emptyset, \\[.5em]
0 & \ \ \mbox{otherwise}.
\end{array}\right.
\eeq

\section{Lieb-Robinson bound from the discrete convolution}

The nested structure of (\ref{iter3}) might be exploited to invoke the discrete convolution in reducing the infinite HK series to a closed form of the LR bound \cite{FFG}. The basic strategy is to conceive a device in the discrete convolution to take advantage of the compromise between the exponential and the power-law decays as reflected by
\beq
\|\widetilde{h}^m_{\ZlrI}\|\leqslant\frac{{\sf c}\cdot e^{-m}}{[1+\textrm{diam}(Z^{lr})]^\eta},
\label{exp_pow}
\eeq
where Eqs.~(\ref{bound1}), (\ref{bound2}), and (\ref{cond1}) have been used and the constant ${\sf c}$ properly encapsulates all the relevant parameters to ensure the validity of (\ref{exp_pow}) for any $Z^{lr},m$.

Next, take a fixed $Z^{lr}_0$ for example, one shall always be able to find a way to divide and assign a realization of $L_0$ and $R_0$ that satisfies $L_0\cup R_0=Z^{lr}_0$ and simultaneously fulfills the definitions of the nearby two $D$-functions that involve $Z^{lr}_0$. Since eventually $L_0$ and $R_0$ will be treated as two independent $Z^{lr}_0$, this step might lead to looser bound but should not cause any violation of the inequality. $L_0$ and $R_0$ can have overlaps. Now assume the existence of such a division, (\ref{iter3}) can then be rewritten as follows,
\begin{align}
&\|[A^\ell_I(t),\mathcal{S}(t)B\mathcal{S}^\dagger(t)]\| \nonumber \\
&\leqslant\|[A^\ell_I(t),B]\|+2\|A^\ell_I(t)\|\sum_{L_0,R_0}\sum_{m_{0L},m_{0R}}\mathcal{D}_i(t;L_0,m_{0L})\times\frac{2{\sf c}\cdot e^{-\frac{m_{0L}+m_{0R}}{2}}}{[1+\textrm{dmax}(L_0,R_0)]^\eta}\mathcal{D}_f(t;R_0,m_{0R})\|B\|\int\limits^t_0d\tau_0 \nonumber \\
&+2\|A^\ell_I(t)\|\sum_{L_0,R_0}\sum_{m_{0L},m_{0R}}\mathcal{D}_i(t;L_0,m_{0L})\frac{2{\sf c}\cdot e^{-\frac{m_{0L}+m_{0R}}{2}}}{[1+\textrm{dmax}(L_0,R_0)]^\eta}\times\sum_{L_1,R_1}\sum_{m_{1L},m_{1R}}\mathcal{D}(t;R_0,m_{0R};L_1,m_{1L}) \nonumber \\
&\times\frac{2{\sf c}\cdot e^{-\frac{m_{1L}+m_{1R}}{2}}}{[1+\textrm{dmax}(L_1,R_1)]^\eta}\mathcal{D}_f(t;R_1,m_{1R})\|B\|\int\limits^t_0d\tau_0\int\limits^{\tau_0}_0 d\tau_1 \nonumber \\
&+2\|A^\ell_I(t)\|\sum_{L_0,R_0}\sum_{m_{0L},m_{0R}}\mathcal{D}_i(t;L_0,m_{0L})\frac{2{\sf c}\cdot e^{-\frac{m_{0L}+m_{0R}}{2}}}{[1+\textrm{dmax}(L_0,R_0)]^\eta}\times\sum_{L_1,R_1}\sum_{m_{1L},m_{1R}}\mathcal{D}(t;R_0,m_{0R};L_1,m_{1L}) \nonumber \\
&\times\frac{2{\sf c}\cdot e^{-\frac{m_{1L}+m_{1R}}{2}}}{[1+\textrm{dmax}(L_1,R_1)]^\eta}\sum_{L_2,R_2}\sum_{m_{2L},m_{2R}}\mathcal{D}(t;R_1,m_{1R};L_2,m_{2L}) \nonumber \\
&\times\frac{2{\sf c}\cdot e^{-\frac{m_{2L}+m_{2R}}{2}}}{[1+\textrm{dmax}(L_2,R_2)]^\eta}\mathcal{D}_f(t;R_2,m_{2R})\|B\|\int\limits^t_0d\tau_0\int\limits^{\tau_0}_0 d\tau_1\int\limits^{\tau_1}_0 d\tau_2+\cdots,
\label{iter4}
\end{align}
where in the final step, the summations have been relaxed to the independent subsets of $L_i$ and $R_i$, whose maximal distance is parametrized by $\textrm{dmax}(L_i,R_i)$. Analogous extensions also apply to the summations of the integer exponents.

Following \cite{FFG}, we perform the summation over the integer exponents first. Take the first term for example. If $\textrm{dmin}(R_0,L_1)\leqslant 2R(t)$, then
\begin{align}
\sum^\infty_{m_{0R}=0}&\sum^\infty_{m_{1L}=0}e^{-\frac{m_{0R}}{2}}\cdot\mathcal{D}(t;R_0,m_{0R};L_1,m_{1L})\cdot e^{-\frac{m_{1L}}{2}}=\sum^\infty_{m_{0R}=0}\sum^\infty_{m_{1L}=0}e^{-\frac{m_{0R}}{2}}\cdot e^{-\frac{m_{1L}}{2}}=\left(\frac{\sqrt{e}}{\sqrt{e}-1}\right)^2=\delta^2.
\end{align}
If instead $\textrm{dmin}(R_0,L_1)>2R(t)$, then
\begin{align}
\sum^\infty_{m_{0R}=0}&\sum^\infty_{m_{1L}=0}e^{-\frac{m_{0R}}{2}}\cdot\mathcal{D}(t;R_0,m_{0R};L_1,m_{1L})\cdot e^{-\frac{m_{1L}}{2}} \nonumber \\
&=\sum^\infty_{m_{0R}=\left\lceil\frac{\textrm{dmin}(R_0,L_1)-2R(t)}{2\chi}\right\rceil}e^{-\frac{m_{0R}}{2}}\cdot\sum^\infty_{m_{1L}=0}e^{-\frac{m_{1L}}{2}}\leqslant\delta^2e^{-[\textrm{dmin}(R_0,L_1)-2R(t)]/4\chi}.
\end{align}
In general,
\begin{align}
\sum^\infty_{m_{iR}=0}&\sum^\infty_{m_{i+1,L}=0}e^{-\frac{m_{iR}}{2}}\cdot\mathcal{D}(t;R_i,m_{iR};L_{i+1},m_{i+1,L})\cdot e^{-\frac{m_{i+1,L}}{2}} \nonumber \\
&=\mathcal{K}(t;R_i;L_{i+1})=\left\{
\begin{array}{cl}
\delta^2 & \ \ \mbox{if}\ \ \ \textrm{dmin}(R_i,L_{i+1})\leqslant2R(t), \\[.5em]
\delta^2e^{-[\textrm{dmin}(R_i,L_{i+1})-2R(t)]/4\chi} & \ \ \mbox{if}\ \ \ \textrm{dmin}(R_i,L_{i+1})>2R(t).
\end{array}\right.
\label{Kdef}
\end{align}
According to this result, (\ref{iter4}) simplifies to
\begin{align}
&\sum^\infty_{\ell=0}\|[A^\ell_I(t),\mathcal{S}(t)B\mathcal{S}^\dagger(t)]\|\leqslant\sum^\infty_{\ell=0}\|[A^\ell_I(t),B]\|+2c_A\sum_{L_0,R_0}\mathcal{K}_i(t;L_0)\frac{2{\sf c}}{[1+\textrm{dmax}(L_0,R_0)]^\eta}\mathcal{K}_f(t;R_0)\|B\|\int\limits^t_0d\tau_0 \nonumber \\
&+2c_A\sum_{L_0,R_0}\mathcal{K}_i(t;L_0)\frac{2{\sf c}}{[1+\textrm{dmax}(L_0,R_0)]^\eta}\sum_{L_1,R_1}\mathcal{K}(t;R_0;L_1)\times\frac{2{\sf c}}{[1+\textrm{dmax}(L_1,R_1)]^\eta}\mathcal{K}_f(t;R_1)\|B\|\int\limits^t_0d\tau_0\int\limits^{\tau_0}_0 d\tau_1 \nonumber \\
&+2c_A\sum_{L_0,R_0}\mathcal{K}_i(t;L_0)\frac{2{\sf c}}{[1+\textrm{dmax}(L_0,R_0)]^\eta}\sum_{L_1,R_1}\mathcal{K}(t;R_0;L_1)\frac{2{\sf c}}{[1+\textrm{dmax}(L_1,R_1)]^\eta} \nonumber \\
&\times\sum_{L_2,R_2}\mathcal{K}(t;R_1;L_2)\frac{2{\sf c}}{[1+\textrm{dmax}(L_2,R_2)]^\eta}\mathcal{K}_f(t;R_2)\|B\|\int\limits^t_0d\tau_0\int\limits^{\tau_0}_0 d\tau_1\int\limits^{\tau_1}_0 d\tau_2+\cdots,
\label{iter5}
\end{align}
where the summation of the index $\ell$ has been first extended and then performed over the initial and final configurations and the constant $c_A$ is defined by
\beq
c_A=\|A\||X|p^{-1}_0s'(1+e).
\eeq

To make further progress, three key features of the constituent functions in (\ref{iter5}) need to be explored. First, it can be noticed from (\ref{Kdef}) that in the region of short distance, function $\mathcal{K}$ exhibits a flat plateau. Second, once beyond the dynamical length scale, $\mathcal{K}$ decays exponentially. Third, the connecting function between the neighboring $\mathcal{K}$ functions decays, in contrast, as a power law of the separation as is inherited from (\ref{exp_pow}). Therefore, the observations that (1) function $\mathcal{K}$ possesses a two-stage structure and (2) the hybridization of the exponential versus the power-law decaying functions play a vital role in the discrete convolution of the HK series.

We now proceed in three successive steps to obtain or better estimate the LR bound.
\begin{enumerate}
\item[(1)] Qualitatively, it is not hard to estimate the convolution of an exponentially decaying function with a power law, therefore
\begin{align}
\sum_{L_i}&\mathcal{K}(t;R_{i-1};L_i)\frac{1}{[1+\textrm{dmax}(L_i,R_i)]^\eta} \nonumber \\
&\leqslant\mathcal{F}(t;R_{i-1};R_i)\approx\left\{
\begin{array}{cl}
\lambda_\chi & \ \ \mbox{if}\ \ \ \textrm{dmin}(R_{i-1},R_{i})<wR(t), \\[.5em]
\lambda_\chi\frac{[wR(t)]^\eta}{[1+\textrm{dmin}(R_{i-1},R_i)]^\eta} & \ \ \mbox{if}\ \ \ \textrm{dmin}(R_{i-1},R_{i})\geqslant wR(t).
\end{array}\right.
\end{align}
This result is based on several observations \cite{FFG}. First, for those $\textrm{dmin}(R_{i-1},R_{i})<wR(t)$ where $w$ is an adjustable coefficient of order $1$, function $\mathcal{K}$ can be simply replaced by its maximum value $\delta^2$, then a dimensional analysis leads to the rough estimate that,
\beq
\sum_L\frac{1}{[1+\textrm{dmax}(L,R)]^\eta}\lessapprox\lambda\chi^{D-\eta},
\eeq
where $\lambda$ is of order $1$ and $D$ is the spatial dimension. Thus,
\beq
\lambda_\chi\coloneqq\lambda\delta^2\chi^{D-\eta}.
\eeq
Second, when $\textrm{dmin}(R_{i-1},R_{i})\geqslant wR(t)$, it is easy to understand that the major contributions of the summation shall come from those $L_i$'s that are closer to $R_{i-1}$ so that the exponential suppression in $\mathcal{K}$ can be compensated to some degree and the resulting convolution shall then be approximated by an overall power law at long distance between $R_{i-1}$ and $R_{i}$. Third, the continuity condition of $\mathcal{F}$ at $\textrm{dmin}(R_{i-1},R_{i})\approx wR(t)$ finally sets the various forms of the coefficients. Be aware that the two-stage structure of function $\mathcal{F}$ is a direct reflection of the two-stage structure of function $\mathcal{K}$. Accordingly, (\ref{iter5}) reduces to
\begin{align}
&\sum^\infty_{\ell=0}\|[A^\ell_I(t),\mathcal{S}(t)B\mathcal{S}^\dagger(t)]\|\leqslant\sum^\infty_{\ell=0}\|[A^\ell_I(t),B]\|+2c_A\sum_{R_0}2{\sf c}\mathcal{F}_i(t;R_0)\mathcal{K}_f(t;R_0)\|B\|\int\limits^t_0d\tau_0 \nonumber \\
&+2c_A\sum_{R_0}2{\sf c}\mathcal{F}_i(t;R_0)\sum_{R_1}2{\sf c}\mathcal{F}(t;R_0;R_1)\mathcal{K}_f(t;R_1)\|B\|\int\limits^t_0d\tau_0\int\limits^{\tau_0}_0 d\tau_1 \nonumber \\
&+2c_A\sum_{R_0}2{\sf c}\mathcal{F}_i(t;R_0)\sum_{R_1}2{\sf c}\mathcal{F}(t;R_0;R_1)\times\sum_{R_2}2{\sf c}\mathcal{F}(t;R_1;R_2)\mathcal{K}_f(t;R_2)\|B\|\int\limits^t_0d\tau_0\int\limits^{\tau_0}_0 d\tau_1\int\limits^{\tau_1}_0 d\tau_2+\cdots.
\label{iter6}
\end{align}
\item[(2)] The treatment of the discrete convolution of the $\mathcal{F}$ functions relies on the cousin of the reproducing condition (\ref{cond2}), which basically states that
\beq
\sum_{R_{i}}\widetilde{\mathcal{F}}(t;R_{i-1};R_i)\widetilde{\mathcal{F}}(t;R_i;R_{i+1})\leqslant g[R(t)]^D\widetilde{\mathcal{F}}(t;R_{i-1};R_{i+1}),
\eeq
where to keep the dimension correct, we redefine
\beq
\widetilde{\mathcal{F}}(t;R_i;R_{j})\coloneqq\frac{1}{\lambda_\chi}\mathcal{F}(t;R_i;R_{j}).
\eeq 
\item[(3)] Finally, by noticing that generically power-law functions $\widetilde{\mathcal{F}}$ decay more slowly than the exponential functions $\mathcal{K}$, one might be able to perform the following replacement,
\beq
\mathcal{K}_f(t;R)\ \ \ \Longrightarrow\ \ \ \widetilde{\mathcal{F}}_f(t;R).
\eeq
\end{enumerate}

Combine all these simplifications and do the integrals of times, one can readily obtain from (\ref{iter6}) the following,
\begin{align}
\sum^\infty_{\ell=0}\|[A^\ell_I(t),\mathcal{S}(t)B\mathcal{S}^\dagger(t)]\|&\leqslant\sum^\infty_{\ell=0}\|[A^\ell_I(t),B]\|+2c_A\|B\|\cdot\frac{2{\sf c}\lambda_\chi g[R(t)]^D}{\left\{\textrm{dist}(X,Y)/[wR(t)]\right\}^\eta}\cdot t \nonumber \\
&+2c_A\|B\|\cdot\frac{\left\{2{\sf c}\lambda_\chi g[R(t)]^D\right\}^2}{\left\{\textrm{dist}(X,Y)/[wR(t)]\right\}^\eta}\cdot \left(\frac{1}{2}t^2\right) \nonumber \\
&+2c_A\|B\|\cdot\frac{\left\{2{\sf c}\lambda_\chi g[R(t)]^D\right\}^3}{\left\{\textrm{dist}(X,Y)/[wR(t)]\right\}^\eta}\cdot \left(\frac{1}{3!}t^3\right)+\cdots \nonumber \\
&\leqslant\sum^\infty_{\ell=0}\|[A^\ell_I(t),B]\|+2c_A\|B\|w^{-\eta}\cdot\frac{\exp\!\left\{2{\sf c}\lambda_\chi g[R(t)]^D\cdot t\right\}}{\left[\textrm{dist}(X,Y)/R(t)\right]^\eta}.
\label{iter7}
\end{align}
Next, the overall short-range contribution can be derived as follows.
\begin{align}
\sum^\infty_{\ell=0}\|[A^\ell_I(t),B]\|&\approx\sum^\infty_{\ell=\left\lceil\frac{r}{\chi}-vt\right\rceil}\|[A^\ell_I(t),B]\|\leqslant\sum^\infty_{\ell=\left\lceil\frac{r}{\chi}-vt\right\rceil}2\|A^\ell_I(t)\|\|B\|\leqslant\sum^\infty_{\ell=\left\lceil\frac{r}{\chi}-vt\right\rceil}4c_A\|B\|e^{-\ell}=4c_A\|B\|\frac{e}{e-1}e^{vt-r/\chi},
\end{align}
where we have assumed that the sizes of the supports of operators $A,B$ are negligibly small as compared to their separation $r\coloneqq\textrm{dist}(X,Y)$ so that the involved commutators are nonzero only if
\beq
\ell\geqslant \frac{r}{\chi(0)}-\frac{\chi(t)vt}{\chi(0)}\gtrapprox\frac{r}{\chi(t)}-vt.
\eeq
Here, we have also exclusively focused on the spacetime regimes that are close to the light-cone front. Hence, we finally obtain the LR bound through the interaction-picture formulation of a generic $k$-body nonlocal interacting Hamiltonian with power-law decaying strengths,
\begin{align}
\|[A_H&(t),B]\|\leqslant\sum^\infty_{\ell=0}\|[A^\ell_I(t),\mathcal{S}(t)B\mathcal{S}^\dagger(t)]\| \nonumber \\
&\leqslant4c_A\|B\|\frac{e}{e-1}e^{vt-\textrm{dist}(X,Y)/\chi(t)}+2c_A\|B\|w^{-\eta}\cdot\frac{\exp\!\left\{2{\sf c}\lambda_\chi g[R(t)]^D\cdot t\right\}}{\left[\textrm{dist}(X,Y)/R(t)\right]^\eta} \nonumber \\
&=2\|A\||X|\|B\|p^{-1}_0s'(1+e)\left\{\frac{2e}{e-1}e^{vt-\textrm{dist}(X,Y)/\chi(t)}+w^{-\eta}\cdot\frac{\exp\!\left\{2{\sf c}\lambda_\chi g[R(t)]^D\cdot t\right\}}{\left[\textrm{dist}(X,Y)/R(t)\right]^\eta}\right\}.
\label{iter8}
\end{align}
Up to some prefactors, the refined LR bound (\ref{iter8}) resembles that obtained by \cite{FFG} and thus shares the same form of the standard LR bound in long-range power-law systems first derived by Hastings and Koma \cite{HastingsKoma}. The essential improvement resides in the renormalization of the various contents dressed by the introduced dynamical length scale, which somehow can be anticipated from scratch. Be cautious, rather than directly working with the bare lattice Hamiltonian as was done in Refs.~\cite{FFG,HastingsKoma}, we have reformulated a modified HK series using the integral-of-motion representation, which is suitable in the context of MBL.

\bibliography{cMBLOTOCSupp}